\documentclass[aps,nofootinbib,amsfonts,superscriptaddress]{revtex4}
\usepackage{amsfonts,amssymb,amscd,amsmath}
\usepackage{graphicx}
\usepackage{subfigure}
\usepackage{bm}
\graphicspath{{figures/}}

\newcommand{\rr}{\mbox{\boldmath $r$}}
\newcommand{\rb}{\mbox{\boldmath $b$}}

\newcommand{\dd}{\, \mathrm{d}}

\begin{document}

\begin{flushright}
LU TP 16-03\\
April 2016
\vskip1cm
\end{flushright}

\title{Nuclear effects in Drell-Yan pair production in high-energy $pA$ collisions}

\author{Eduardo Basso}
\email{eduardo.basso@thep.lu.se}
\affiliation{Instituto de F\'{\i}sica, Universidade Federal do Rio de Janeiro, 
Caixa Postal 68528, Rio de Janeiro, RJ 21941-972, Brazil}

\author{Victor P.  Goncalves}
\email{victor.goncalves@thep.lu.se}
\affiliation{Department of Astronomy and Theoretical Physics, Lund University, 
SE-223 62 Lund, Sweden} 
\affiliation{High and Medium Energy Group, Instituto de F\'{\i}sica e 
Matem\'atica, Universidade Federal de Pelotas, Pelotas, RS, 96010-900, Brazil} 

\author{Michal Krelina}
\email{michal.krelina@fjfi.cvut.cz}
\affiliation{Czech Technical University in Prague, FNSPE, B\v rehov\'a 7, 11519 
Prague, Czech Republic}

\author{Jan Nemchik}
\email{nemcik@saske.sk}
\affiliation{Czech Technical University in Prague, FNSPE, B\v rehov\'a 7, 11519 
Prague, Czech Republic}
\affiliation{Institute of Experimental Physics SAS, Watsonova 47, 04001 Ko\v 
sice, Slovakia}

\author{Roman Pasechnik}
\email{roman.pasechnik@thep.lu.se}
\affiliation{Department of Astronomy and Theoretical Physics, Lund University, 
SE-223 62 Lund, Sweden}

\begin{abstract}
The Drell-Yan (DY) process of dilepton pair production off nuclei is not affected by final state interactions, energy loss or absorption. 
A detailed phenomenological study of this process is thus convenient for investigation of the onset of initial-state effects in 
proton-nucleus ($pA$) collisions. In this paper, we present a comprehensive analysis of the DY process in $pA$ interactions 
at RHIC and LHC energies in the color dipole framework. We analyse several effects affecting the nuclear suppression, 
$R_{pA}<1$, of dilepton pairs, such as the saturation effects, restrictions imposed by energy conservation (the initial-state 
effective energy loss) and the gluon shadowing, as a function of the rapidity, invariant mass of dileptons and their transverse 
momenta $p_T$. In this analysis, we take into account besides the $\gamma^*$ also the $Z^0$ contribution to the production 
cross section, thus extending the predictions to large dilepton invariant masses. Besides the nuclear attenuation of produced dileptons at large energies 
and forward rapidities emerging due to the onset of shadowing effects, we predict a strong suppression at large $p_T$, dilepton invariant 
masses and Feynman $x_F$ caused by the Initial State Interaction effects in kinematic regions where no shadowing is expected. 
The manifestations of nuclear effects are investigated also in terms of the correlation function in azimuthal angle between the dilepton pair and a 
forward pion $\Delta\phi$ for different energies, dilepton rapidites and invariant dilepton masses. We predict that the characteristic 
double-peak structure of the correlation function around $\Delta \phi\simeq \pi$ arises for very forward pions and large-mass dilepton pairs.
\end{abstract}
\maketitle

\section{Introduction}

During the last two decades, a series of theoretical and experimental studies of particle production in heavy ion collisions (HICs) at 
Relativistic Heavy Ion Collider (RHIC) and Large Hadrons Collider (LHC) energies has been performed. These results provided us with various 
sources of information on properties of the hot and dense matter (Quark Gluon Plasma) formed in these collisions. 
Although several issues still remain open, those are mainly related to a description of nuclear effects related to the initial-state 
formation before it interacts with a nuclear target, as well as to the parton propagation in a nuclear medium. 
In this context, the phenomenological studies of hard processes in proton-nucleus ($pA$) collisions can provide us with 
an additional quantitative information about various nuclear effects expected also in HICs. This can help us to disentangle 
between the medium effects of different types and constrain their relative magnitudes and contributions \cite{salgado}. 

A key feature of the Drell-Yan (DY) process is the absence of final state interactions and fragmentation associated 
with an energy loss or absorption phenomena. For this reason, the DY process can be considered as a very clean probe 
for the Initial State Interaction (ISI) effects \cite{peng}. In practice, this process can be used as a convenient tool in studies 
of the Quantum Chromodynamics (QCD) at high energies, in particular, the saturation effects expected to determine the initial conditions 
in hadronic collisions as well as the initial-state energy loss due to the projectile quark propagation in the nuclear medium before 
it experiences a hard scattering.

In the present paper, we study the DY process on nuclear targets at high energies using the color dipole approach 
\cite{k95,bhq97,kst99,krt01,dynuc,rauf,gay,pkp,Basso,Basso_pp}, which is known to give as precise prediction for the DY cross section 
as the Next-to-Leading-Order (NLO) collinear factorization framework and allows to include naturally the coherence effects in nuclear collisions. 
Moreover, the color dipole formalism provides a straightforward generalisation of the DY process description from the proton-proton 
to proton-nucleus collisions and is thus suitable for studies of nuclear effects directly accessing the impact parameter dependence 
of nuclear shadowing and nuclear broadening -- the critical information which is not available in the parton model.

\begin{figure}[t]
\large
\begin{center} 
  \scalebox{0.7}{\includegraphics{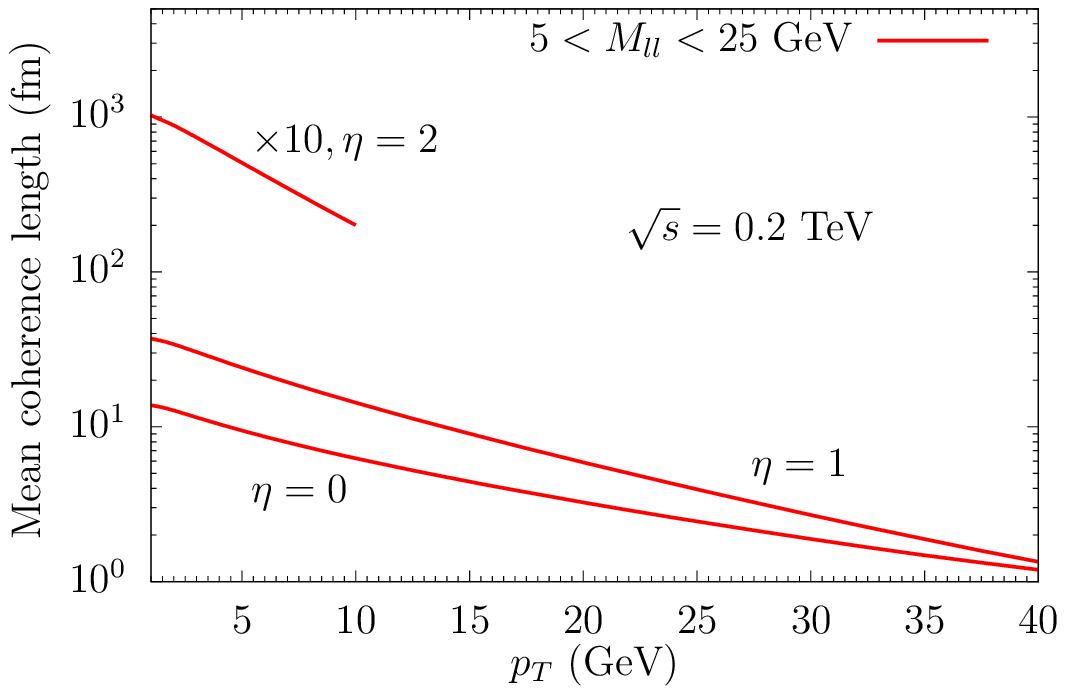}}
  \scalebox{0.7}{\includegraphics{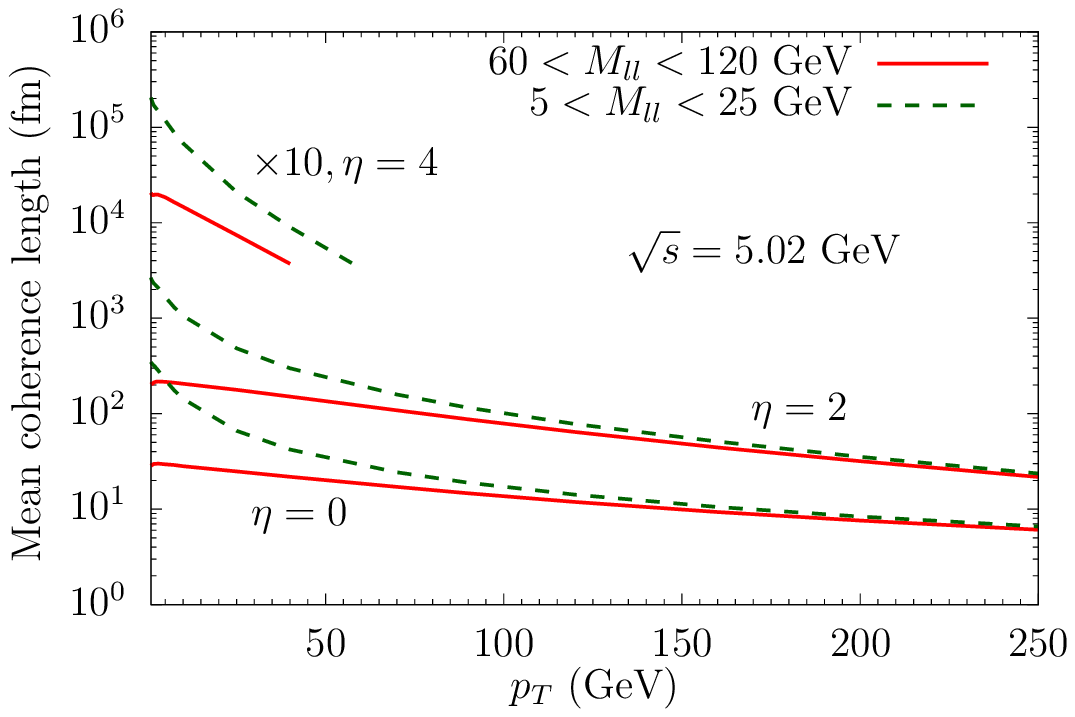}}
  \caption{(Color online) 
  The mean coherence length $l_c$ of the DY reaction 
  in $pA$ collisions at RHIC and LHC energies 
  for different dilepton rapidities and invariant mass ranges.}
  \label{fig:LCL}
\end{center}
\end{figure}

In contrast to the conventional parton model where the dilepton production 
process is typically viewed as the parton annihilation in the center of mass (c.m.) 
frame, in the color dipole approach operating in the target rest frame 
the same process looks as a bremsstrahlung of a $\gamma^*/Z^0$ boson 
off a projectile quark. In $pA$ collisions assuming the high energy limit, 
the projectile quark probes a dense gluonic field in the target 
and the nuclear shadowing leads to a nuclear modification of the
transverse momentum distribution of the DY production cross section. 
The onset of shadowing effects is controlled by the coherence length,
which can be interpreted as the mean lifetime of $\gamma^*/Z^0$-quark
fluctuations, and is given by 
\begin{equation}
\label{eq-cl}
  l_c = \frac{1}{x_2 m_N} 
\frac
{(M_{l\bar{l}}^2 + p_T^2)(1-\alpha)}
{ \alpha(1-\alpha) M_{l\bar{l}}^2 + \alpha^2 m_f^2 + p_T^2} \,,
\end{equation}
where $M_{l\bar{l}}$ is the dilepton invariant mass  and  $p_T$ its transverse momentum. Moreover, $\alpha$ is the fraction 
of the light-cone momentum of the projectile quark carried out by the gauge boson. As demonstrated in Fig. \ref{fig:LCL},  in the RHIC and LHC kinematic regions, the coherence length exceeds the nuclear radius $R_A$, 
$l_c \gtrsim R_A$, which implies that  
the long coherence length (LCL) limit can be safely used 
in practical calculations of the DY cross section in $pA$ collisions. 

Besides the quark 
shadowing effects naturally accounted for 
in the dipole picture, one should also take into account the nuclear effects 
due to multiple rescattering of initial-state projectile partons (ISI effects) in a medium 
before a hard scattering. The latter are important close to the kinematic limits, 
e.g. at large Feynman $x_F\to 1$ and $x_T=2 p_T/\sqrt{s}\to 1$ ($\sqrt{s}$ is the 
collision energy in c.m. frame), due to restrictions imposed by energy conservation. 
In the present paper, we take into account also non-linear QCD effects, 
which are amplified in nuclear collisions and related to multiple scatterings 
of the higher Fock states containing gluons in the dipole-target interactions.
They generate the gluon shadowing effects effective at small Bjorken $x$
in the target and large rapidity values.

\normalsize
In our study, all the basic ingredients for the DY nuclear production 
cross section (such as the dipole cross section parameterisations 
and Parton Distribution Functions (PDFs)) have been determined from other processes. 
Consequently, our predictions are parameter-free and should be considered as
an important test for the onset of distinct nuclear effects. 
Note that the nuclear DY process mediated by a virtual photon has been already 
studied within the color dipole framework by several authors 
(see e.g. Refs.~\cite{dynuc,rauf,gay}). However,
the results of this paper represent a further step 
updating and improving the previous analyses in the literature 
providing new predictions for the transverse momentum, 
dilepton invariant mass and rapidity distributions 
of the nuclear DY production cross section 
at RHIC and LHC energies as well as in comparison to the most recent data. 
Besides, the effects of quantum coherence at large energies including 
the gluon shadowing as a leading-twist shadowing correction as well as 
an additional contribution of the $Z^0$ boson and $\gamma^*/Z^0$ interference 
are incorporated. Moreover, the impact of the effective 
initial state energy loss effects on the DY nuclear production cross section 
is studied for the first time. We also investigate nuclear effects providing 
a detailed analysis of the azimuthal correlation between 
the produced DY pair and a forward pion taking into 
account the $Z^0$ boson contribution 
in addition to virtual photon, 
generalising thus the results presented in Ref.~\cite{stastody}.

This paper is organized as follows. In the Section~\ref{formalism}, 
we present a brief overview of gauge boson production in the color dipole framework. 
Moreover, we discuss in detail the saturation effects, gluon shadowing 
and initial-state energy loss effects included in the analysis. 
Section~\ref{res} is devoted to predictions for the dilepton invariant mass, 
rapidity and transverse momentum distributions of the DY nuclear production 
cross sections in comparison with the available data. 
The onset of various nuclear effects is estimated in the LCL limit 
and the predictions for the nucleus-to-nucleon ratio, 
$R_{pA}=\sigma_{pA}/A\sigma_{pp}\footnote{Here $A$ represents the atomic mass
number of the nuclear target}$, of the DY production 
cross sections are presented. 
The latter can be verified in the future by experiments at RHIC an LHC. 
Furthermore, the azimuthal correlation function between the produced dilepton 
and a pion is evaluated for $pA$ collisions at RHIC and LHC for different 
dilepton invariant masses including the high-mass region. 
Finally, in Section~\ref{conc} we summarise our main conclusions.

%
%
%
%
\section{Drell-Yan process in hadron-nucleus collisions}
\label{formalism}
%
%
%
%

%
%
\subsection{DY nuclear cross section}
%
%

The color dipole formalism is treated in the target rest frame 
where the process of DY pair production can be viewed as a radiation 
of gauge bosons $G^*=\gamma^*/Z^0$  by a projectile quark 
(see e.g. Ref.~\cite{Basso_pp,pkp}). Assuming only the lowest $|qG^*\rangle$ 
Fock component, the cross section for the inclusive gauge boson 
production with invariant mass $M_{l\bar{l}}$ and transverse momentum $p_T$ 
can be expressed in terms of the projectile quark (antiquark) 
densities $q_f$ ($\bar{q}_f$) at momentum fraction $x_q$  
and the quark-nucleus cross section as follows (see e.g. Refs.~\cite{dynuc,Basso_pp}),
%
\begin{eqnarray} 
\label{eq:gb_cs}
\frac{d \sigma (pA\rightarrow G^* X)}{\dd^2 p_T\, d\eta} = 
J(\eta, p_T)\,\frac{x_1}{x_1 + x_2}\,
\sum_f\sum_{\lambda_G=L,T}\,
  \int\limits_{x_1}^1 
\frac{d \alpha}{\alpha^2}\, 
\bigl [
\, q_f(x_q,\mu_F^2) + \bar{q}_{{f}}(x_q,\mu_F^2)  
\bigr ]\,
\frac{d\sigma^f_{\lambda_G} (qA \rightarrow qG^*X)}{d(\ln\alpha)\, d^2p_T}\, ,
\end{eqnarray}
%
where 
%
\begin{equation}
J(\eta, p_T)\equiv \frac{dx_F}{d\eta} = 
\frac{2}{\sqrt{s}} \sqrt{M_{l\bar{l}}^2 + p_T^2}\, \cosh(\eta)
\end{equation}
%
is the Jacobian of transformation between the Feynman variable $x_F = x_1 - x_2$ 
and pseudorapidity $\eta$ of the virtual gauge boson 
$G^*$,  $x_q = x_1/\alpha$, where $\alpha$ is the fraction 
of the light-cone momentum of the projectile quark carried out by the gauge boson, 
and $\mu_F^2=p_T^2+(1-x_1)M_{l\bar{l}}^2$ is the factorization scale in quark PDFs. 
As in Ref. \cite{Basso_pp} we take $\mu_F\simeq M_{l\bar{l}}$, for simplicity. 

The transverse momentum distribution in Eq.~({\ref{eq:gb_cs}) of the gauge boson $G^*$ 
bremsstrahlung in quark-nucleus interactions
can be obtained by a generalization of the well-known formulas 
for the photon bremsstrahlung from Refs.~\cite{dynuc,rauf,kst99}.
Then the corresponding differential cross section for a given incoming 
quark of flavour $f$ reads,
%
\begin{eqnarray}
\label{ptdistcc}
\frac{d\sigma^f_{T,L} (qA \rightarrow qG^*X)}{d(\ln\alpha)\,d^2p_T} 
& = & 
\frac{1}{(2\pi)^2}\, \sum_\text{quark pol.}
\int d^2\rho_1\,d^2\rho_2\,
\exp\bigl [i\,{\bf p}_T \cdot ({\bm\rho}_1 - {\bm\rho}_2)\bigr ]\, 
\Psi^{\cal{V-A}}_{T,L}(\alpha,{\bm\rho}_1,m_f)\,
\Psi^{\cal{V-A},*}_{T,L}(\alpha,{\bm\rho}_2,m_f) \nonumber \\
&\times & 
\frac{1}{2}\bigl [ \sigma_{q\bar{q}}^A(\alpha {\bm\rho}_1,x_2) + 
\sigma_{q\bar{q}}^A(\alpha {\bm\rho}_2,x_2) - 
\sigma_{q\bar{q}}^A(\alpha|{\bm\rho}_1- {\bm\rho}_2|,x_2)\bigr]\,, 
\end{eqnarray}
%
where $x_2 = x_1 - x_F$ and ${\bm\rho}_{1,2}$ are 
the quark-$G^*$ transverse separations in the total radiation amplitude 
and its conjugated counterpart, respectively. 
Assuming that the projectile quark is unpolarized, 
the vector $\Psi^{\cal{V}}$ and axial-vector $\Psi^{\cal{A}}$ wave functions 
in Eq.~(\ref{ptdistcc}) are not correlated such that
%
\begin{eqnarray} 
\label{Psi2}
&    & 
\sum_\text{quark pol.} \Psi^{\cal{V-A}}_{T,L}(\alpha,{\bm\rho}_1,m_f)\, 
\Psi^{\cal{V-A},*}_{T,L}(\alpha,{\bm\rho}_2,m_f)  = 
\nonumber \\ 
& = & 
\Psi^{\cal{V}}_{T,L}(\alpha,{\bm\rho}_1,m_f)\,
\Psi^{\cal{V},*}_{T,L}(\alpha,{\bm\rho}_2,m_f) + 
\Psi^{\cal{A}}_{T,L}(\alpha,{\bm\rho}_1,m_f)\, 
\Psi^{\cal{A},*}_{T,L}(\alpha,{\bm\rho}_2,m_f)\,,
\end{eqnarray}
%
where the averaging over the initial and summation 
over final quark helicities is performed and 
the quark flavour dependence comes 
only via the projectile quark mass $m_f$. 
The corresponding wave functions 
$\Psi_{T,L}^{\cal{V-A}}(\alpha,{\bm\rho})$ can be found in Ref.~\cite{pkp}.

Our goal is to evaluate the DY production cross section 
in $pA$ collisions at high energies and a large mass number $A$
of the nuclear target. 
This regime is characterised by a limitation 
on the maximum phase-space parton density that can be reached 
in the hadron wave function (parton saturation) \cite{hdqcd}. 
The transition between the linear and non-linear regimes 
of QCD dynamics is typically specified by a characteristic 
energy-dependent scale called the saturation scale 
$Q_{s}^2$, where the variable $s$ denotes c.m. energy squared of the collision.
Such saturation effects are expected to be amplified 
in nuclear collisions since the nuclear saturation scale $Q_{s,A}^2$ 
is expected to be enlarged with respect to 
the nucleon one $Q_{s,p}^2$ by rougthly a factor of $A^{1/3}$.

In general, the dipole-nucleus cross section 
$\sigma_{q\bar{q}}^A(\rho,x)$ can be written
in terms of the forward dipole-nucleus scattering amplitude 
${\cal N}^A (\rho, x, \rb)$ as follows,
%
\begin{eqnarray}
\sigma_{q\bar{q}}^A(\bm\rho,x) = 2\,\int d^2\rb\,  {\cal N}^A (\bm\rho, x, \rb)\,.
\label{sigda}
\end{eqnarray}
%
At high energies, the evolution of $\mathcal{N}^A(x,\rr,\rb)$ 
in rapidity $Y = \ln (1/x)$ is given, for example, within 
the Color Glass Condensate (CGC) formalism \cite{CGC}, in terms 
of an infinite hierarchy of equations known as so called 
Balitsky-JIMWLK equations \cite{BAL,CGC}, 
which reduces in the mean field approximation 
to the Balitsky-Kovchegov (BK) equation \cite{BAL,kov}. 
In recent years, several groups have studied 
the solution of the BK equation taking into account 
the running coupling corrections to the evolution kernel. 
However, these analyses have assumed the translational 
invariance approximation, which implies that
$\mathcal{N}^A(\rho, x,\rb) = \mathcal{N}^A(\rho,x)\,S(\rb)$ and 
$\sigma_{q\bar{q}}^A(\bm\rho, x, \rb) = \sigma_0\,\mathcal{N}(\rho, x)$, 
where $\mathcal{N}(\rho, x)$ is a partial dipole amplitude on a nucleon, 
and $\sigma_0$ is the normalization of the dipole cross section 
fitted to the data. Basically, they disregard the impact parameter dependence. 
Unfortunately, the impact-parameter dependent numerical solutions 
of the BK equation are very difficult to obtain \cite{stastobk}. 
Moreover, the choice of the impact-parameter profile 
of the dipole amplitude entails intrinsically nonperturbative physics, 
which is beyond the QCD weak coupling approach of the BK equation. 
In what follows, we explore an alternative path and 
employ the available phenomenological models, which explicitly 
incorporate an expected $b$-dependence of the scattering amplitude.
%
%

%
%
\subsection{Models for the dipole cross section}
%
%

As in our previous studies 
\cite{erike_ea2,vmprc,vic_erike,babi,Diego,Diego2}, 
we work in the LCL limit and employ the model 
initially proposed in Ref.~\cite{kopeliovich-lcl} 
which includes the impact parameter dependence 
in the dipole-nucleus amplitude and describes 
the experimental data on the nuclear structure function 
(for more details, see Ref.~\cite{armesto,erike_ea2}).
In particular, this model enables us to incorporate 
the shadowing effects via a simple
eikonalization of the standard dipole-nucleon 
cross section $\sigma_{q\bar q}(\bm\rho,x)$ such that the forward 
dipole-nucleus amplitude in Eq.~(\ref{sigda}) is given by
%
\begin{eqnarray}
{\cal N}^A (\bm\rho,x,\rb) = 
1-\exp\left(-\frac{1}{2}\,T_A(\rb)\,\sigma_{q\bar{q}}(\bm\rho,x)\right) \,,
\label{Na}
\end{eqnarray}
%
where $T_A(\rb)$ is the nuclear profile (thickness) 
function, which is normalized to the mass number $A$
and reads
%
\begin{eqnarray}
T_A(\rb) =
\int_{-\infty}^{\infty} \rho_A(\rb,z) dz\, .
\end{eqnarray}
%
Here $\rho_A(\rb,z)$ represents 
the nuclear density function defined at the impact parameter $\rb$
and the longitudinal coordinate $z$. In our calculations we
used realistic parametrizations of $\rho_A(\rb,z)$ from
Ref.~\cite{saxon}.
The eikonal formula (\ref{Na}) based upon the Glauber-Gribov 
formalism \cite{gribov} 
resums the multiple elastic rescattering diagrams 
of the $q\bar{q}$ dipole in a nucleus in the high-energy limit.
The eikonalisation procedure is justified in the LCL regime where 
the transverse separation $\rho$ of partons in the multiparton 
Fock state of the photon 
is frozen during propagation through the nuclear matter and
becomes an eigenvalue of the scattering matrix. 

For the numerical analysis of the nuclear DY observables, 
we need to specify a reliable parametrisation for 
the dipole-proton cross section. 
In recent years, several groups have constructed a number 
of viable phenomenological models based on saturation physics 
and fits to the HERA and RHIC data (see e.g. 
Refs.~\cite{GBW,iim,kkt,dhj,Goncalves:2006yt,buw,kmw,agbs,Soyez2007,bgbk,kt,ipsatnewfit,amirs}). 

As in our previous study of the DY process in $pp$ collisions
\cite{Basso_pp},
in order to estimate theoretical uncertainty in our analysis, 
in what follows, 
we consider several phenomenological models for the dipole cross 
section $\sigma_{q\bar{q}}$ which take into account the DGLAP evolution 
as well as the saturation effects. 

The first one is the model proposed in Ref.~\cite{bgbk}, 
where the dipole cross section is given by
%
\begin{equation}
 \sigma_{q\bar{q}}(\bm\rho, x) = \sigma_0\,
\left[1-\exp\left( - \frac{\pi^2}{\sigma_0\,N_c}\,\rho^2\,
 \alpha_s(\mu^2)\,xg(x, \mu^2)\right)\right]\,,
 \label{bgbk}
\end{equation}
%
where $N_c=3$ is the number of colors, $\alpha_s(\mu^2)$ 
is the strong coupling constant at $\mu$ scale, 
which is related to the dipole size 
$\rho$ as $\mu^2=C/\rho^2 + \mu_0^2$ 
with $C$, $\mu_0$ and $\sigma_0$ parameters fitted 
to the HERA data. Moreover, in this model 
the gluon density evolves according to DGLAP equation \cite{dglap} 
accounting for gluon splittings only,
%
\begin{equation}
\frac{\partial xg(x,\mu^2)}{\partial \ln \mu^2 } = 
\frac{\alpha_s(\mu^2)}{2\pi} \int_x^1 dz\,  
P_{gg}(z) \frac{x}{z} g\Big(\frac{x}{z}, \mu^2\Big)\,, 
\label{dglap}
\end{equation}
%
where the gluon density at initial scale $\mu_0^2$ is parametrized as
\cite{bgbk}
%
\begin{equation}
xg(x,\mu_0^2) = A_g x^{-\lambda_g} (1-x)^{5.6}\,.
\end{equation}
%
The set of best fit values of the model parameters reads: 
$A_g = 1.2$, $\lambda_g = 0.28$, $\mu_0^2 = 0.52$ GeV$^{2}$, 
$C = 0.26$ and $\sigma_0 = 23$ mb. 
In what follows we denote by BGBK the predictions 
for the DY observables obtained using Eq.~(\ref{bgbk}) as an input 
in calculations of the dipole-nucleus scattering amplitude. 

The model proposed in Ref.~\cite{bgbk} was generalised 
in Ref.~\cite{kmw} in order to take into 
account the impact parameter dependence of 
the dipole-proton cross section and to describe 
the exclusive observables at HERA. In this model, 
the corresponding dipole-proton cross section is given by
%
\begin{eqnarray}
\sigma_{q\bar{q}}(\bm\rho, x) = 
2\,\int d^2b_p\,\left[1-\exp\left(- \frac{\pi^2}{2N_c}\,
\rho^2\,\alpha_s(\mu^2)\, xg(x,\mu^2)T_G({\bf b}_p)\right)\right]
\label{ipsat}
\end{eqnarray} 
%
with the DGLAP evolution of the gluon distribution given 
by Eq.~(\ref{dglap}). The Gaussian impact parameter dependence 
is given by 
$T_G({\bf b_p})=(1/2\pi B_G)\,\exp(-b_p^2/2 B_G)$, 
where $B_G$ is a free parameter extracted from the $t$-dependence of 
the exclusive electron-proton ($ep$) data. 
The parameters of this model were updated in 
Ref.~\cite{ipsatnewfit} by fitting to the recent high precision 
HERA data \cite{heradata} providing the following values: 
$A_g = 2.373$, $\lambda_g = 0.052$, $\mu_0^2 = 1.428$ GeV$^{2}$, 
$B_G = 4.0$ GeV$^{2}$ and $C = 4.0$. 
Hereafter, we will denote as IP-SAT the resulting predictions 
obtained using Eq.~(\ref{ipsat}) 
as an input in calculations of ${\cal N}^A$, Eq.~(\ref{Na}).

For comparison with the previous results existing in the literature, 
we also consider the Golec-Biernat-Wusthoff (GBW) model \citep{GBW}
based upon a simplified saturated form
%
\begin{equation}
\label{gbw}
\sigma_{q\bar{q}}(\bm\rho,x) = 
\sigma_0\,\left(1 - e^{-\frac{\rho^2Q_s^2(x)}{4}}\right)
\end{equation}
%
with the saturation scale
%
\begin{equation}
Q_s^2(x) = Q_0^2\left( \frac{x_0}{x} \right)^\lambda \,,
\label{satsca1}
\end{equation}
%
where the model parameters $Q_0^2 = 1$ GeV${}^2$, 
$x_0 = 4.01 \times 10^{-5}$, $\lambda = 0.277$ and $\sigma_0 = 29$ mb 
were obtained from the fit to the DIS data accounting for a contribution 
of the charm quark. 

Finally, we also consider the running coupling solution 
of the BK equation for the partial dipole 
amplitude obtained in the Ref.~\cite{bkrunning} 
using the GBW model as an initial condition such that 
$\sigma_{q\bar{q}}^p(\bm\rho,x) = \sigma_0\,{\cal{N}}^p(\bm\rho,x)$ where 
the normalisation $\sigma_0$ is fitted to the HERA data. 

%
%
\subsection{Gluon shadowing corrections}
%
%

In the LHC energy range
the eikonal formula for the
LCL regime, Eq.~(\ref{Na}), is not exact.
Besides the lowest $|qG^*\rangle$ 
Fock state, where $G^*=\gamma^*/Z^0$,
one should include also the higher Fock
components containing gluons, e.g. $|qG^*\,g\rangle$, $|qG^*\,gg\rangle$, etc. 
They cause an additional suppression known as the gluon shadowing (GS). 
Such high LHC energies allow so to activate the coherence effects 
also for these gluon fluctuations, which
are heavier and consequently have a shorter coherence length than
lowest Fock component $|qG^*\rangle$.
The corresponding suppression factor $R_G$, as the ratio 
of the gluon densities in nuclei and nucleon,
was derived 
in Ref.~\cite{kopeliovich-gs}
using the Green function technique
through the calculation of 
the inelastic correction $\Delta\sigma_{tot}(q\bar{q}g)$ to the total
cross section $\sigma_{tot}^{\gamma^*\,A}$, related to the creation
of a $|q\bar{q}\,g\rangle$ intermediate Fock state
%
\begin{eqnarray}
R_G(x,Q^2,\rb) \equiv \frac{x g_A(x,Q^2,\rb)}{A\cdot x g_p(x,Q^2)} 
\approx 1 - \frac{\Delta\sigma_{tot}(q\bar{q}g)}{\sigma_{tot}^{\gamma^*A}} \,.
\end{eqnarray}
%
GS corrections are included in calculations replacing 
$\sigma_{q \bar q}^N(\bm\rho, x) \rightarrow 
\sigma_{q \bar q}^N(\bm\rho, x)\,R_G(x,Q^2,\rb)$.
They lead to additional nuclear suppression
in production of DY pairs at small Bjorken
$x=x_2$ in the target. In Fig.~\ref{fig:rg} (left panel) we present our results for the $x$ dependence of  the ratio $R_G(x,Q^2,\rb)$ for different vales of the impact parameter $b$. As expected, the magnitude of the shadowing corrections decreases at large values of $b$. In the right panel we present our predictions for the $b$-integrated nuclear ratio $R_G(x,Q^2)$ for different values of the hard scale $Q^2$. This figure shows a not very strong onset of GS, which was
confirmed by the NLO global analyses of
DIS data \cite{nlo-dis}. A weak $Q^2$ dependence
of GS demonstrates that GS is a leading twist effect,
with $R_G(x,Q^2)$ approaching unity only very slowly
(logarithmically) as $Q^2\rightarrow\infty$.

\begin{figure}[h!]
\large
\begin{center} 
\scalebox{0.7}{\includegraphics{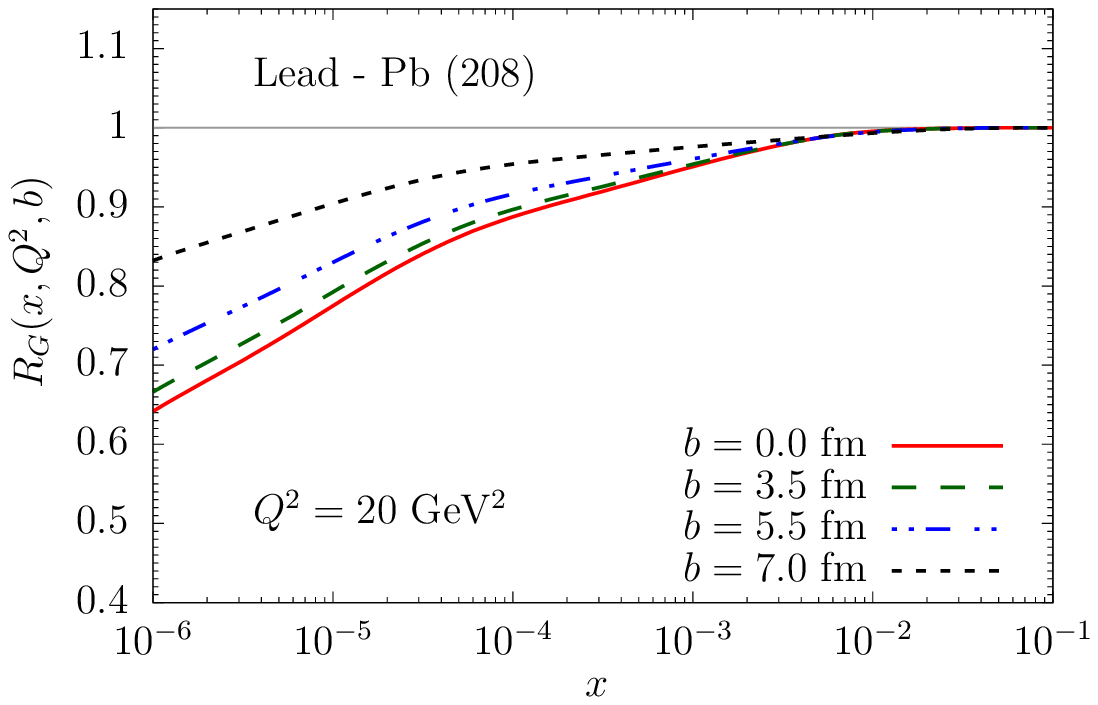}}
\scalebox{0.7}{\includegraphics{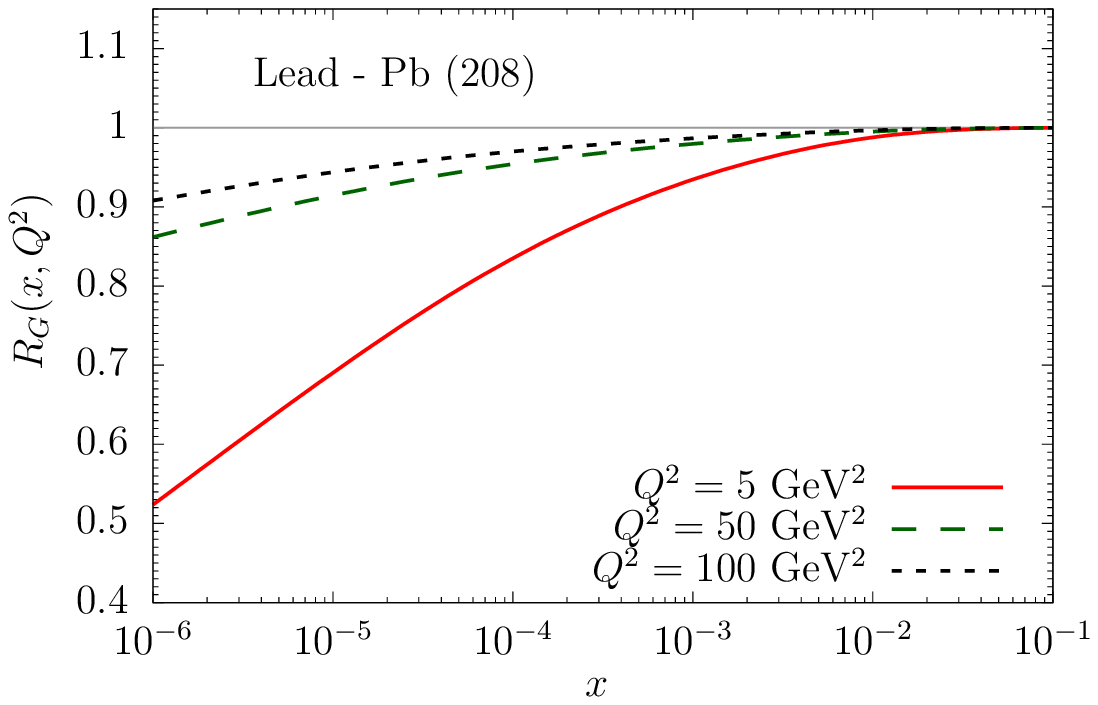}}
\caption{(Color online) Left panel: The $x$-dependence of the ratio 
$R_G(x,Q^2,\rb)$ for different values of the impact parameter. Right panel: The $x$-dependence of the $b$--integrated ratio 
$R_G(x,Q^2)$ for distinct values of the hard scale $Q^2$.}
\label{fig:rg}
\end{center}
\end{figure}
\normalsize

%
%
\subsection{Effective energy loss}
%
%

The effective initial-state energy loss (ISI effects) 
is expected to suppress noticeably the nuclear cross 
section when reaching the kinematical limits, 
\[ 
x_L = \frac{2p_L}{\sqrt{s}}\rightarrow1\,, 
\qquad x_T = \frac{2p_T}{\sqrt{s}}\rightarrow1 \,.
\]
Correspondingly, a proper variable which controls 
this effect is $\xi = \sqrt{x_L^2 + x_T^2}$. The magnitude of suppression 
was evaluated in Ref.~\cite{kopeliovich-isi}. 
It was found within the Glauber approximation 
that each interaction in the nucleus 
leads to a suppression factor $S(\xi)\approx 1-\xi$. 
Summing up over the multiple initial state interactions in a $pA$ collision 
at impact parameter $b$, one arrives at a nuclear ISI-modified PDF
%
\begin{equation}
\label{eq-ISI}
q_{f}(x,Q^2) \Rightarrow q_{f}^A(x,Q^2,b) = 
C_v \, q_{f}(x,Q^2)\,
\frac{e^{-\xi \sigma_{\rm eff}T_A(b)}-e^{-\sigma_{\rm eff}T_A(b)}}
{(1-\xi)(1-e^{-\sigma_{\rm eff}T_A(b)})} \,.
\end{equation}
%
Here, $\sigma_{\rm eff}=20$~mb is the effective hadronic 
cross section controlling the multiple interactions. 
The normalisation factor $C_v$ is fixed by the Gottfried 
sum rule (for more details, see Ref.~\cite{kopeliovich-isi}). 
It was found that such an additional nuclear suppression 
emerging due to the ISI effects represents an energy 
independent feature common for all known reactions 
experimentally studied so far, with any leading particle 
(hadrons, Drell-Yan dileptons, charmonium, etc). 
In particular, such a suppression was indicated at midrapidity, 
$y=0$, and at large $p_T$ by the PHENIX data \cite{phenix-isi-dAu} 
on $\pi^0$ production in central $dAu$ collisions and 
on direct photon production in central $AuAu$ collisions 
\cite{phenix-isi-AuAu}, where no shadowing is expected
since the corresponding Bjorken $x=x_2$ in the target is large.
Besides large $p_T$-values,
the same mechanism of nuclear attenuation is effective also 
at forward rapidities (large Feynman $x_F$), where we expect 
a much stronger onset of nuclear suppression 
as was demonstrated by the BRAHMS and STAR data \cite{rhic-isi-forw}.
In our case, we predict that the ISI effects induce a
significant suppression 
of the DY nuclear cross section at large dilepton $p_T$,
dilepton invariant mass and at forward rapidities as one can see
in the next Section.

%
%
%
\section{Results}
\label{res}
%
%
%

In what follows, we present our predictions for the DY pair production 
cross section in the process $pA\rightarrow \gamma^*/Z^0 \rightarrow l \bar l$ 
obtained within the color dipole formalism 
and taking into account the medium effects discussed in the previous 
Section. 
Following Ref.~\cite{GBW}, we use the quark mass values to be 
$m_u = m_d = m_s = 0.14$ GeV, $m_c = 1.4$ GeV and $m_b = 4.5$ GeV. 
Moreover, we take the factorisation scale $\mu_F$ defined above 
to be equal to the dilepton invariant mass, $M_{l\bar{l}}$, 
and employ the CT10 NLO parametrisation 
for the projectile quark PDFs \cite{ct10} 
(both sea and valence quarks are included). 
As was demonstrated in Refs.~\cite{Basso_pp,Basso:2015lua}, 
there is a little sensitivity of DY predictions on 
PDF parameterisation in $pp$ collisions at high energies 
so we do not vary the projectile quark PDFs.
\begin{figure}[h!]
\large
\begin{center}
\scalebox{0.7}{\includegraphics{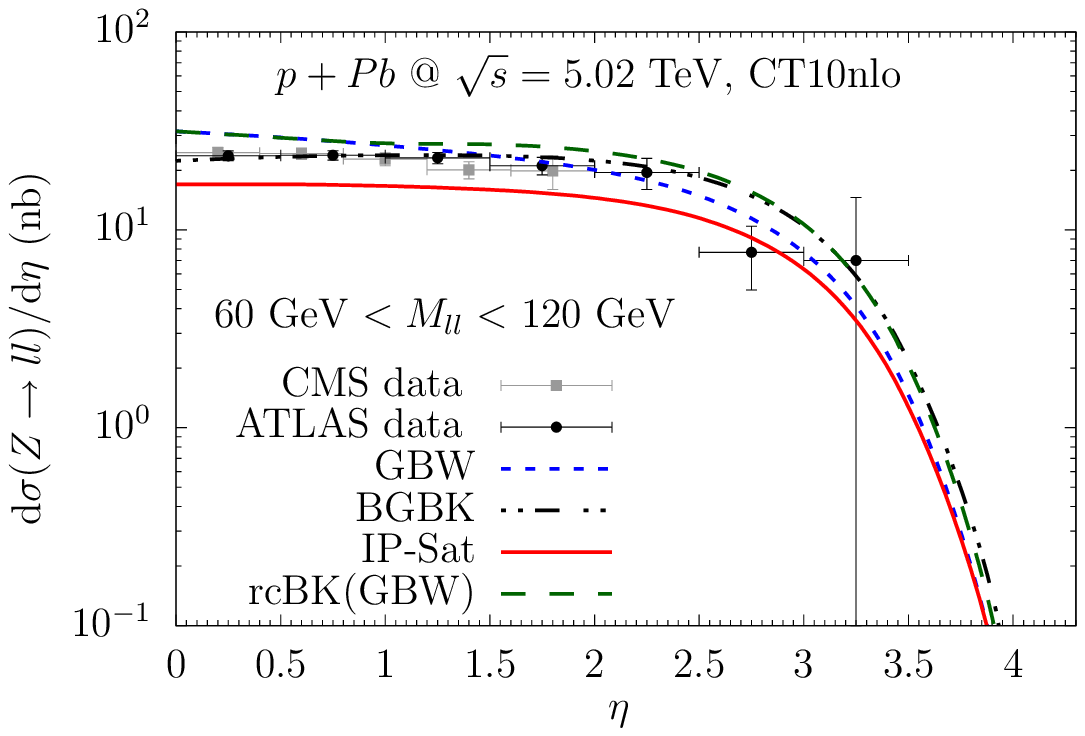}}
\scalebox{0.7}{\includegraphics{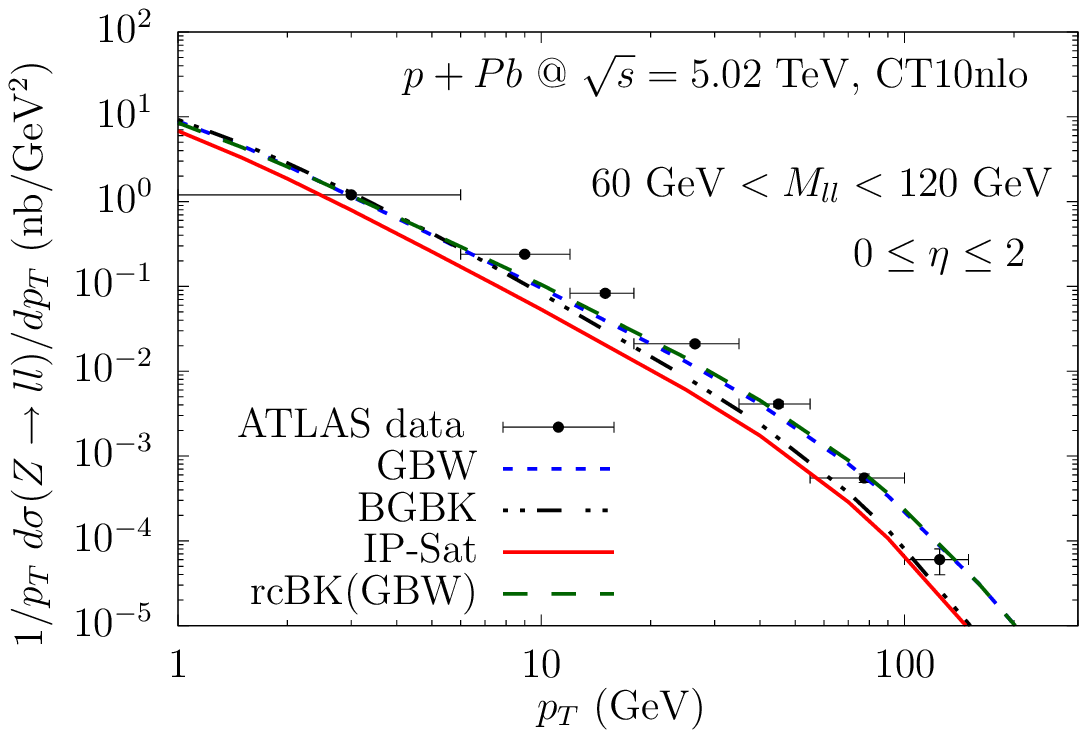}} \\
\scalebox{0.7}{\includegraphics{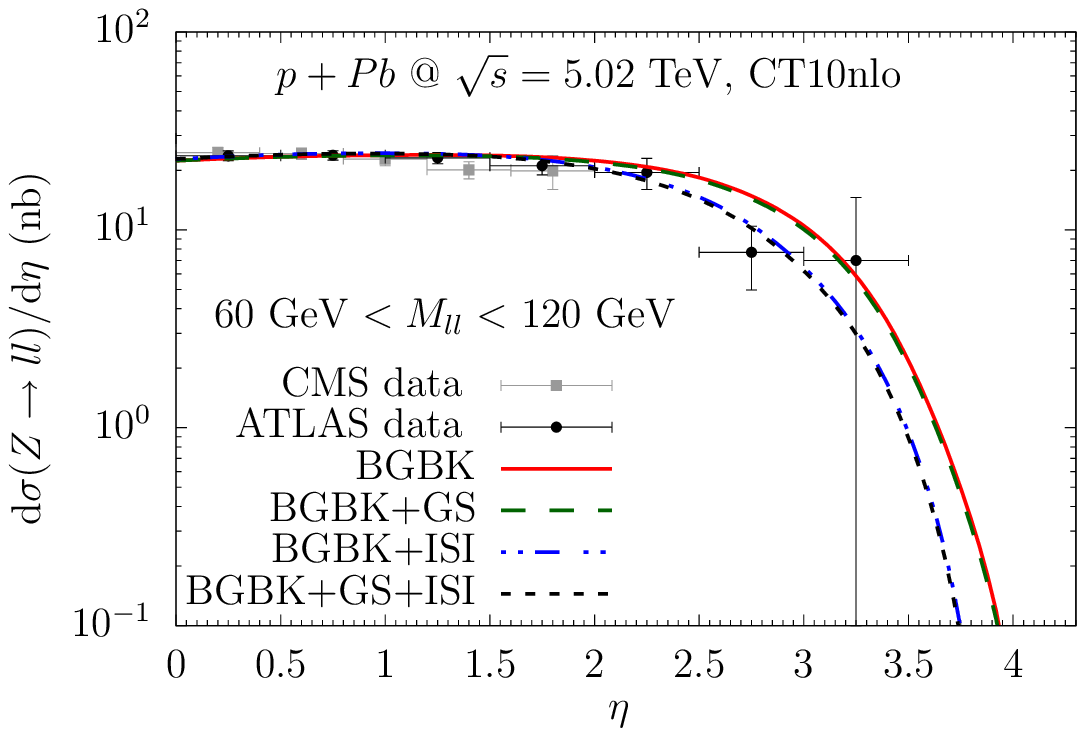}}
\scalebox{0.7}{\includegraphics{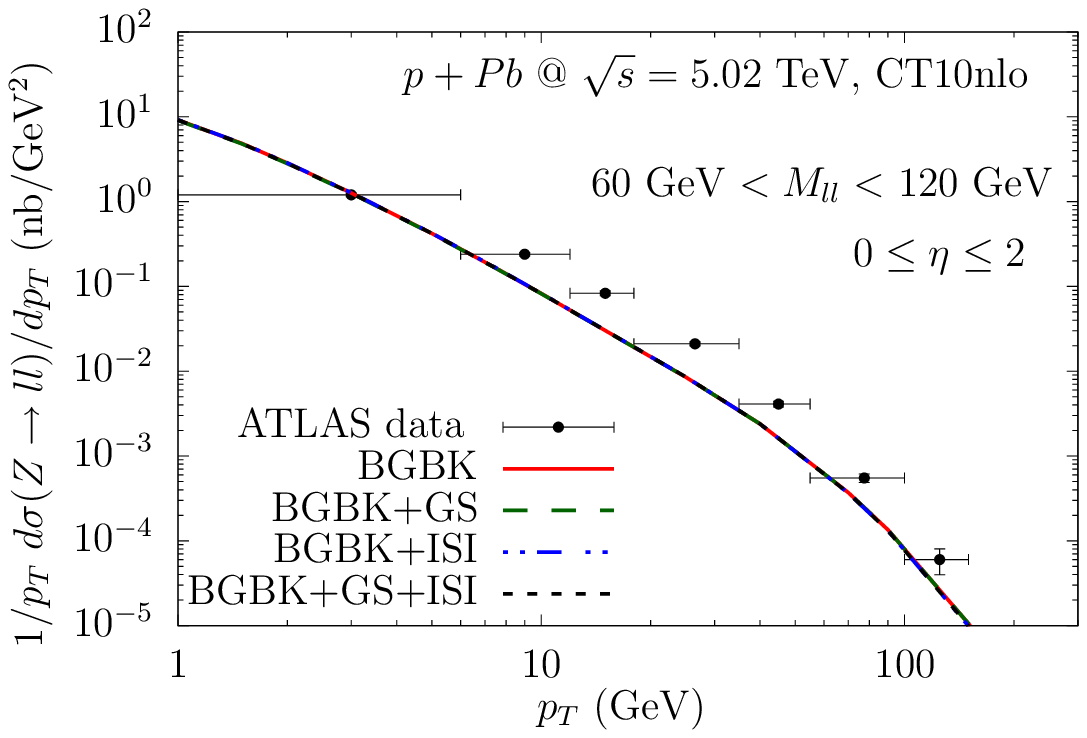}}
\caption{(Color online) The dipole model predictions for the DY 
nuclear cross sections at large dilepton invariant masses 
compared to the recent experimental data from ATLAS and CMS
experiments \cite{cms_data_pA,atlas_data_pA}
at c.m. collision energy $\sqrt{s}=5.02$ TeV.
The predictions obtained 
for several parameterisations of the dipole cross section 
described in the text are shown in the top panels while
the effects of the gluon shadowing and the initial-state energy loss 
are demonstrated in the bottom panels.}
\label{fig:data}
\end{center}
\end{figure}
\normalsize

In Fig.~\ref{fig:data} we compare our predictions for the DY nuclear 
cross section with available LHC data \cite{cms_data_pA,atlas_data_pA} 
for large invariant dilepton masses, $60 < M_{l\bar{l}} < 120$ GeV,
 taking into account the saturation effects. 
In the top panels, we test the predictions of 
various models for the dipole cross section comparing them 
with the experimental data for the rapidity and 
transverse momentum distributions of the DY
production cross sections in $pA$ collisions. 
As was already verified in Ref.~\cite{Basso_pp} 
for DY production in $pp$ collisions, 
the dipole approach works fairly well in description of 
the current experimental data at high energies. 
In particular, the BGBK model provides a consistent prediction 
describing the data on the rapidity distribution 
quite well in the full kinematical range. 
In the bottom panels of Fig.~\ref{fig:data}, we took the BGBK model 
and considered the impact of gluon shadowing corrections
as well as the initial-state effective energy loss (ISI effects),
Eq.~(\ref{eq-ISI}). 
In the range of large dilepton invariant masses 
concerned, the gluon shadowing corrections are rather small
since the corresponding Bjorken $x=x_2$ in the target becomes large.
On the other hand, the ISI effects significantly 
modify the behaviour of the rapidity distribution 
at large $\eta > 2$. Unfortunately, the current data are not able 
at this moment to verify the predicted strong onset of
ISI effects due to large error bars. 
In the case of the transverse momentum distribution 
for large invariant masses and $0 \le \eta \le 2$, 
the impact of both the gluon shadowing 
and the ISI effects is negligible.
\begin{figure}[h!]
\large
\begin{center}
\scalebox{1.0}{\includegraphics{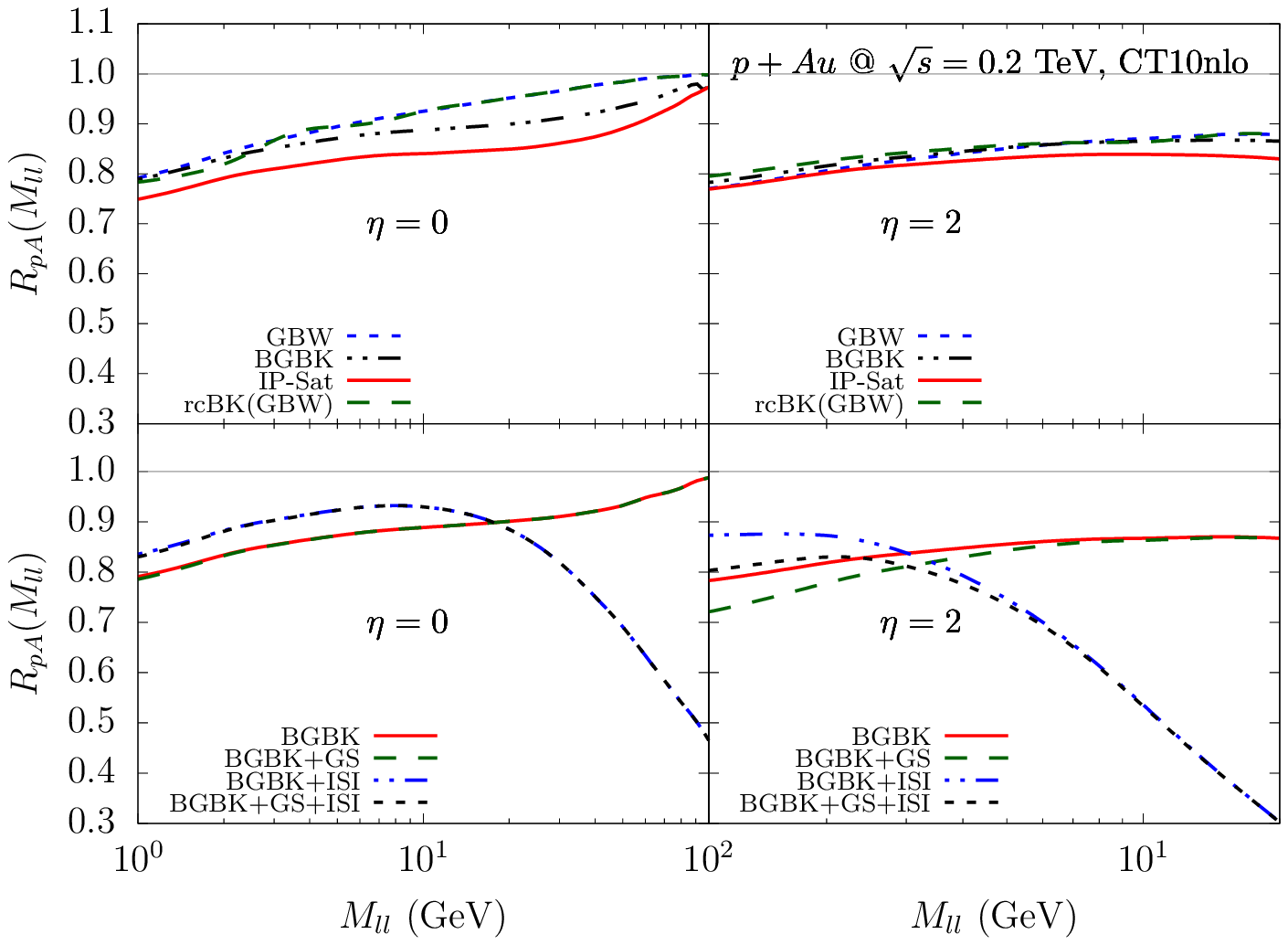}} 
\caption{(Color online) 
The dilepton invariant mass dependence of the nucleus-to-nucleon ratio, 
$R_{pA} = \sigma^{\rm DY}_{pA}/(A \cdot \sigma^{\rm DY}_{pp})$,
of the DY production cross sections for 
c.m. energy $\sqrt{s}=0.2$ TeV corresponding to RHIC experiments.}
\label{fig:mass_rhic}
\end{center}
\end{figure}
\normalsize
\begin{figure}[h!]
\large
\begin{center}
\scalebox{1.0}{\includegraphics{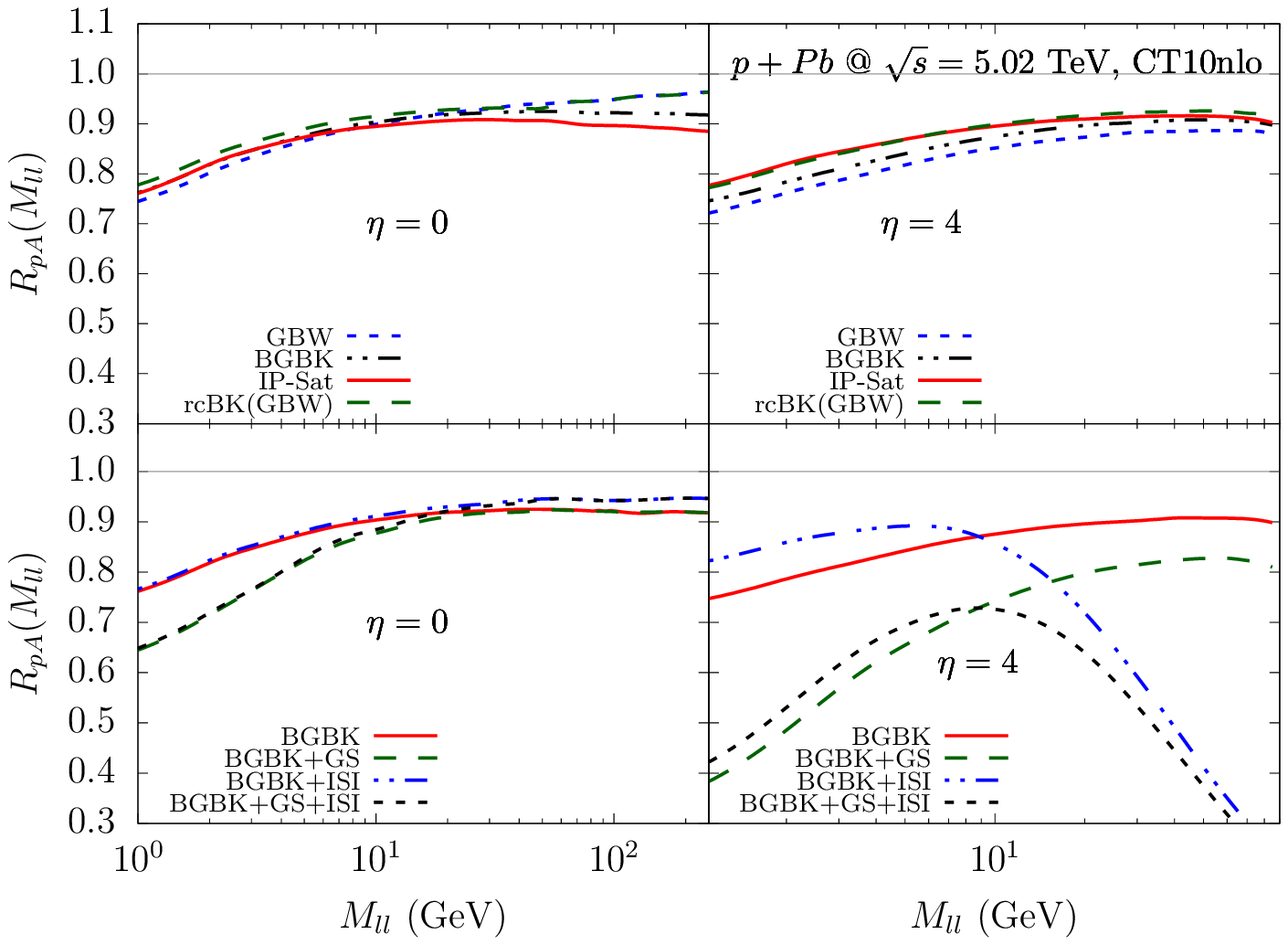}} 
\caption{(Color online) The dilepton invariant mass dependence of the 
nucleus-to-nucleon ratio, 
$R_{pA} = \sigma^{\rm DY}_{pA}/(A \cdot \sigma^{\rm DY}_{pp})$, 
of the DY production cross sections for 
c.m. enegy $\sqrt{s}=5.02$ TeV corresponding to LHC experiments.}
\label{fig:mass_lhc}
\end{center}
\end{figure}
\normalsize

In order to quantify the impact of the nuclear effects, 
in what follows, we estimate the invariant mass, rapidity 
and transverse momentum dependence 
of the nucleus-to-nucleon ratio
of the DY production cross sections (nuclear modification factor), 
$R_{pA} = \sigma^{\rm DY}_{pA}/(A \cdot \sigma^{\rm DY}_{pp})$, 
considering the DY process at RHIC ($\sqrt{s}=0.2$ TeV) and LHC 
($\sqrt{s}=5.02$ TeV) energies. 
The color dipole predictions for the DY production cross section 
in $pp$ collisions have been discussed in detail 
in Ref.~\cite{Basso_pp}. For consistency, 
the numerator and denominator of the nuclear modification factor
are evaluated within the same model for the dipole 
cross section as an input. 

In Fig.~\ref{fig:mass_rhic} we present our predictions 
for the dilepton invariant mass dependence of the ratio 
$R_{pA}(M_{l\bar{l}})$ at RHIC considering both central and 
forward rapidities. 
In the top panels, we show that the dipole model predictions 
are almost insensitive to the parameterisations used to treat 
the dipole-proton interactions. The magnitude of the saturation effects 
decreases at large dilepton invariant masses and increases at forward rapidities. 
Such a behaviour is expected, since at smaller $M_{l\bar{l}}$ and 
at larger $\eta$ one probes smaller values of the Bjorken-$x_2$ variable
in the target.
In the bottom panels of Fig.~\ref{fig:mass_rhic}, 
we present the predictions taking into account also
the GS corrections and ISI effects. 
As was mentioned above we predict a weak
onset of GS corrections 
at central rapidities whereas GS leads to
a significant suppression in the forward region. 
Besides, as expected, the impact of GS effects decreases 
with $M_{l\bar{l}}$ due to rise of the Bjorken $x_2$-values. 
In contrast to that, the ISI effects become effective causing
a strong nuclear suppression
at large $M_{l\bar{l}}$ and/or $\eta$. 
This behaviour is also well understood since large 
dilepton invariant masses and/or rapidities correspond
to large Feynman $x_F$ leading to a stronger onset
of ISI effects as follows from Eq.~(\ref{eq-ISI}).
A similar behaviour has been predicted for the LHC energy
range as is shown in Fig.~\ref{fig:mass_lhc} where the impact 
of saturation and GS effects is even more pronounced.
\begin{figure}[h!]
\large
\begin{center}
\scalebox{1.0}{\includegraphics{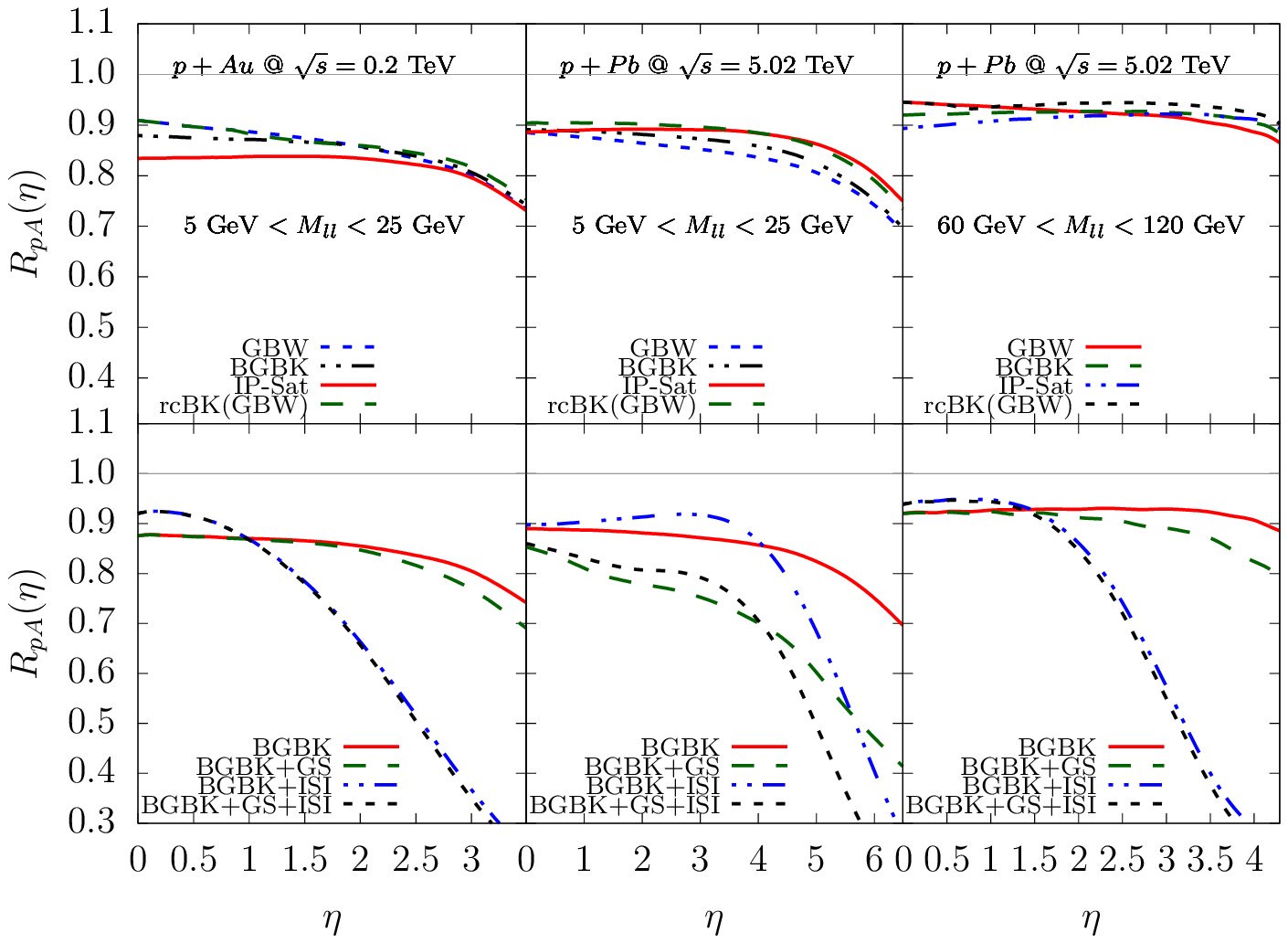}} 
\caption{(Color online) 
The pseudorapidity dependence of the nucleus-to-nucleon ratio, $R_{pA}(\eta)$, 
of the DY production cross sections at RHIC and LHC energies for two ranges
($5 < M_{l\bar{l}} < 25$ GeV)  and ($60 < M_{l\bar{l}} < 120$ GeV) of dilepton invariant mass.}
\label{fig:rapidity}
\end{center}
\end{figure}
\normalsize

In Fig.~\ref{fig:rapidity} we present our predictions 
for rapidity dependence of the nucleus-to-nucleon ratio, $R_{pA}(\eta)$,
of the DY production cross sections at RHIC and LHC energies 
considering two ranges, ($5 < M_{l\bar{l}} < 25$ GeV)  
and ($60 < M_{l\bar{l}} < 120$ GeV), of dilepton invariant mass. 
We would like to emphasize that the onset
of saturation effects reduces $R_{pA}(\eta)$ at large rapidities 
and have a larger impact in the small invariant mass range. 
For large invariant masses, we predict a reduction 
of $\approx 10 \%$ in the $R_{pPb}$ ratio at LHC energy. 
At RHIC energy we predict a weak onset of GS effects
even at large $\eta > 3$.
In contrast to RHIC energy range, 
at the LHC the GS effects lead to a significant additional suppression,
modifying thus the ratio $R_{pPb}$ especially at small dilepton invariant masses
and large rapidity values. 
On the other hand, the onset of the ISI effects 
is rather strong for both RHIC and LHC kinematic regions,
and becomes even stronger at forward rapidities for both 
invariant mass ranges. This makes the phenomenological
studies of the rapidity dependence of $R_{pA}$ ideal for constraining 
such effects. 
\begin{figure}[h!]
\large
\begin{center}
\scalebox{1.0}{\includegraphics{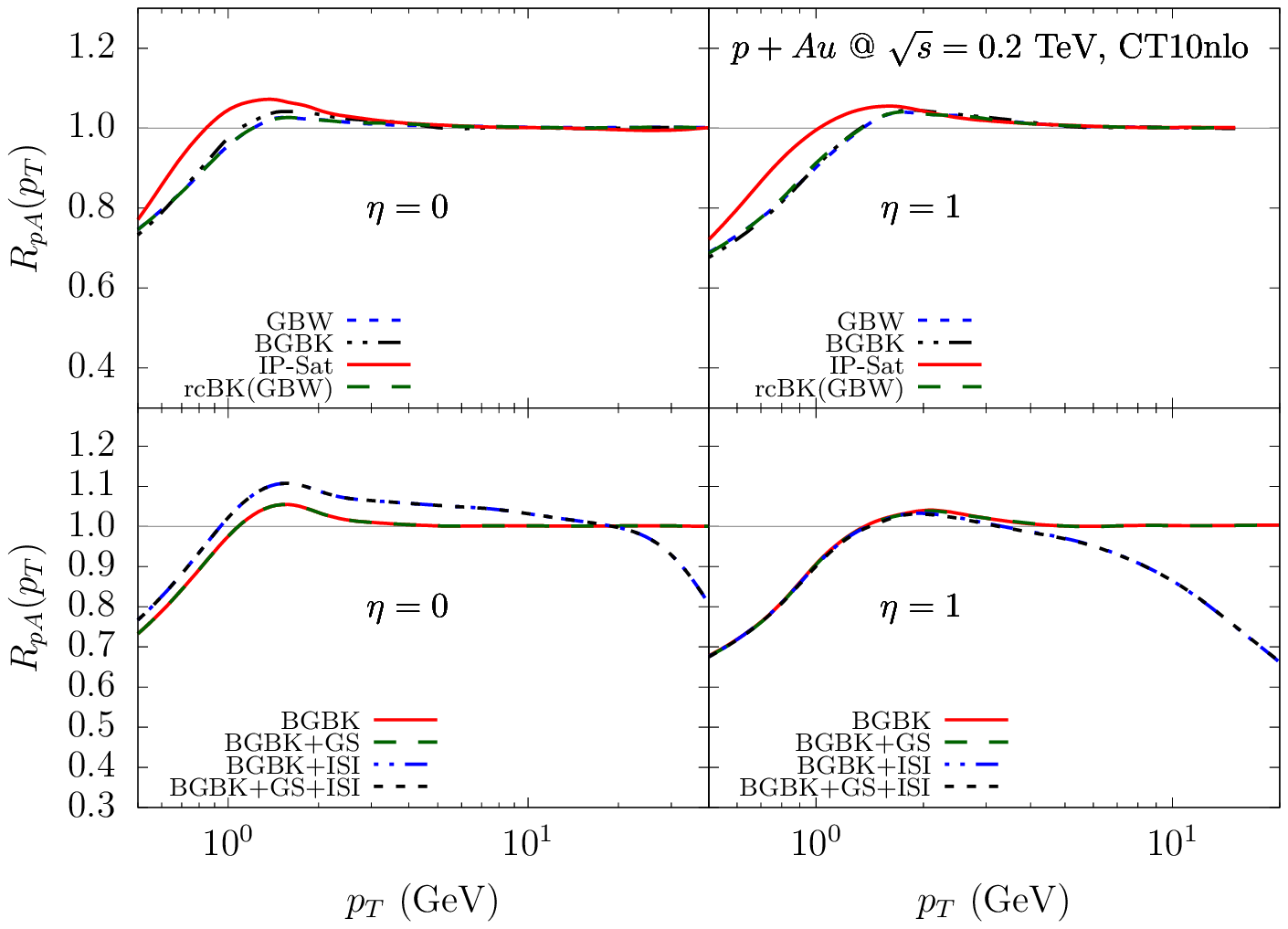}} 
\caption{(Color online) 
The transverse momentum dependence of the nucleus-to-nucleon ratio 
of the DY production cross sections, $R_{pA}(p_T)$,
for the dilepton invariant mass 
range $5 < M_{l\bar{l}} < 25$ GeV at $\sqrt{s}=0.2$ TeV and $\eta=0,1$.}
\label{fig:pt_rhic}
\end{center}
\end{figure}
\normalsize
\begin{figure}[h!]
\large
\begin{center}
\scalebox{1.0}{\includegraphics{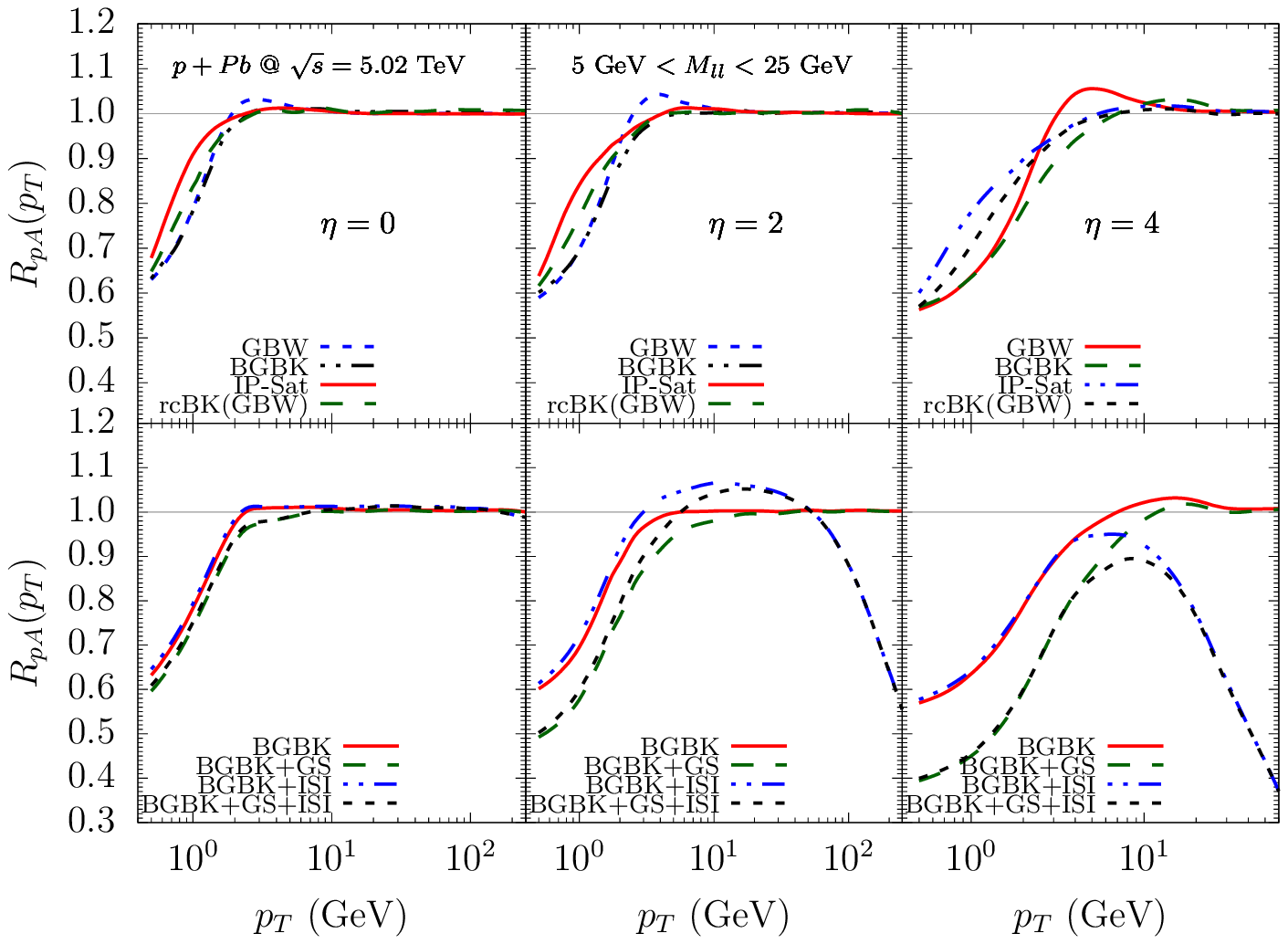}} 
\caption{(Color online) 
The transverse momentum dependence of the nucleus-to-nucleon ratio
of the DY production cross sections, $R_{pA}(p_T)$,
for the dilepton invariant mass 
range $5 < M_{l\bar{l}} < 25$ GeV at $\sqrt{s}=5.02$ TeV and $\eta=0,2,4$.}
\label{fig:pt_lhc_small}
\end{center}
\end{figure}
\normalsize

Fig.~\ref{fig:pt_rhic} shows our predictions for the transverse momentum dependence 
of  the nuclear modification factor, $R_{pA}(p_T)$,
for the invariant mass range 
$5 < M_{l\bar{l}} < 25$ GeV at RHIC c.m. energy $\sqrt{s}=0.2$ TeV 
and two distinct pseudorapidity values $\eta=0$ 
and $\eta=1$. 
At large transverse momenta, the role 
of the saturation effects is negligibly small 
and can be important only at small $p_T \le 2$ GeV. 
Similarly, the GS effects are almost irrelevant at RHIC energies.
However, Fig.~\ref{fig:pt_rhic} clearly demonstrates
a strong onset of ISI effects causing a significant suppression at large $p_T$,
where no coherence effects are expected. In accordance with Eq.~(\ref{eq-ISI})
and in comparison with $\eta=0$,
we predict stronger ISI effects at forward rapidities as is depicted
in Fig.~\ref{fig:pt_rhic} for $\eta=1$. Due to a significant elimination
of coherence effects the study of the DY process at large $p_T$ in $pA$ collisions
at RHIC is a very convenient tool for investigation of net ISI effects.
On the other hand, at LHC energies (see Fig.~\ref{fig:pt_lhc_small}) 
the manifestation of the saturation and GS effects rises at forward rapidities 
and becomes noticeable for $p_T \le 10$ GeV. 
As was already mentioned for RHIC energies, 
the ISI effects cause a significant attenuation 
at large transverse momenta and forward 
rapidities, although no substantial suppression is expected
in the DY process
due to absence of the final state interaction, energy loss or absorption.
For these reasons a study 
of the ratio $R_{pA}(p_T)$ also at the LHC especially at large $p_T$
and at small invariant mass range is very effective to constrain the 
ISI effects.
\begin{figure}[h!]
\large
\begin{center}
\scalebox{1.0}{\includegraphics{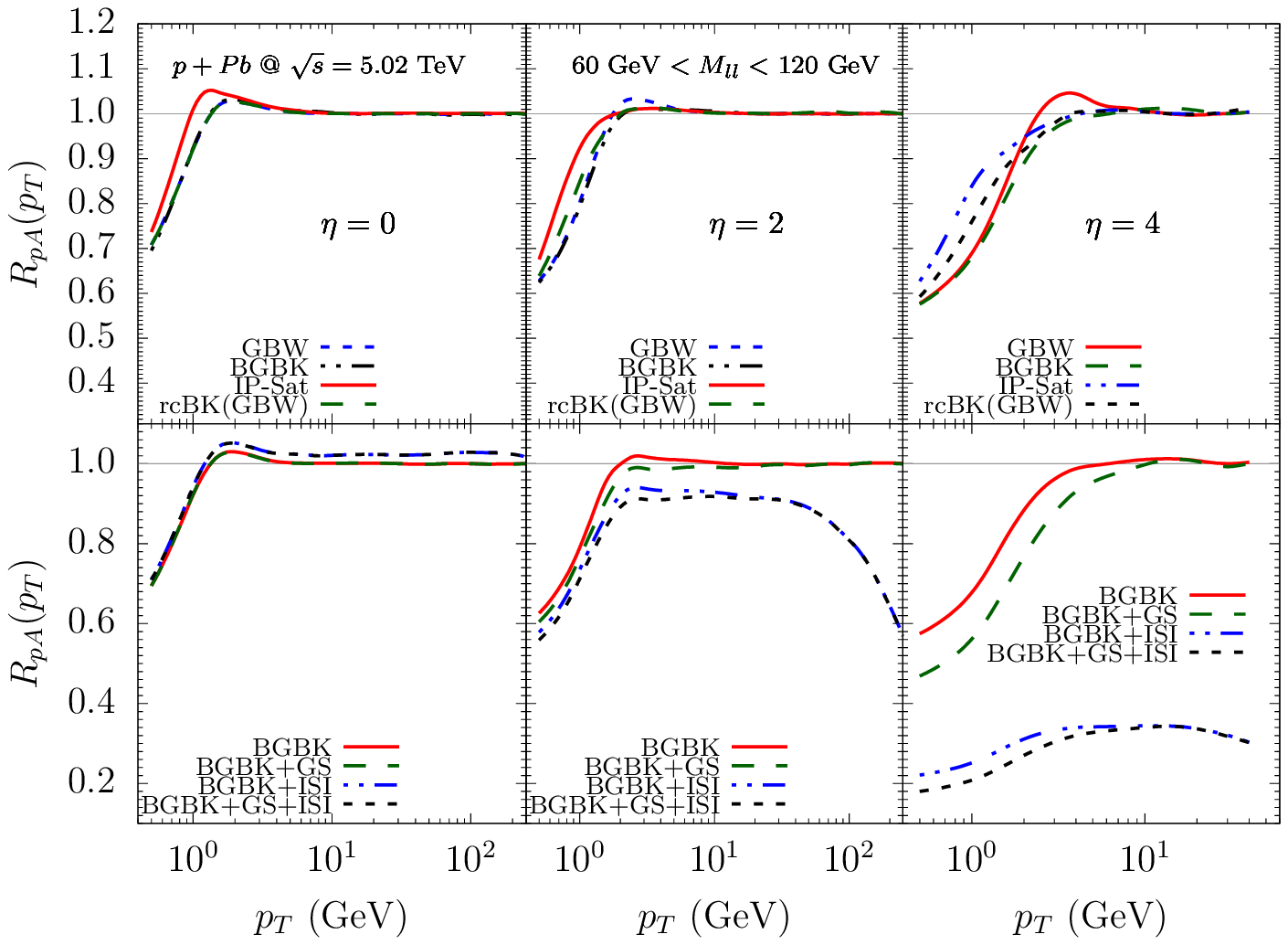}} 
\caption{(Color online) The transverse momentum dependence 
of the nucleus-to-nucleon ratio
of the DY production cross sections, $R_{pA}(p_T)$,
for the dilepton invariant mass 
range $60 < M_{l\bar{l}} < 120$ GeV at $\sqrt{s}=5.02$ TeV and $\eta=0,2,4$.}
\label{fig:pt_lhc_large}
\end{center}
\end{figure}
\normalsize

In order to reduce the contribution of coherence effects (gluon shadowing, CGC) in the LHC kinematic
region one should go to the range of large dilepton invariant masses
as is shown in Fig.~\ref{fig:pt_lhc_large}. Here we
present our predictions for the ratio $R_{pPb}(p_T)$ at the LHC
c.m. collision energy $\sqrt{s}=5.02$ TeV for 
the range $60 < M_{l\bar{l}} < 120$ GeV and several values of $\eta=0,2,4$. 
According to expectations we have found that 
the saturation and GS effects turn out 
to be important only at small $p_T$ and large $\eta$. 
Such an elimination of coherence effects taking
into account larger dilepton invariant masses
causes simultaneously
a stronger onset of ISI effects
as one can seen in Fig.~\ref{fig:pt_lhc_large} 
in comparison with Fig.~\ref{fig:pt_lhc_small}.
For this reason, investigation of net ISI effects
at large $M_{l\bar{l}}$
does not require such high $p_T$- and rapidity values,
what allows to obtain the experimental data of higher
statistics and consequently with smaller error bars.
Fig.~\ref{fig:pt_lhc_large} demonstrates again
a large nuclear suppression in the forward region ($\eta = 4$) over an 
extended range of the dilepton transverse momenta. 
Consequently, such an analysis of the DY nuclear cross section at 
forward rapidities by e.g. the LHCb Collaboration can be 
very useful to probe the ISI effects experimentally.

\begin{figure}[h!]
\large
\begin{center}
\scalebox{0.7}{\includegraphics{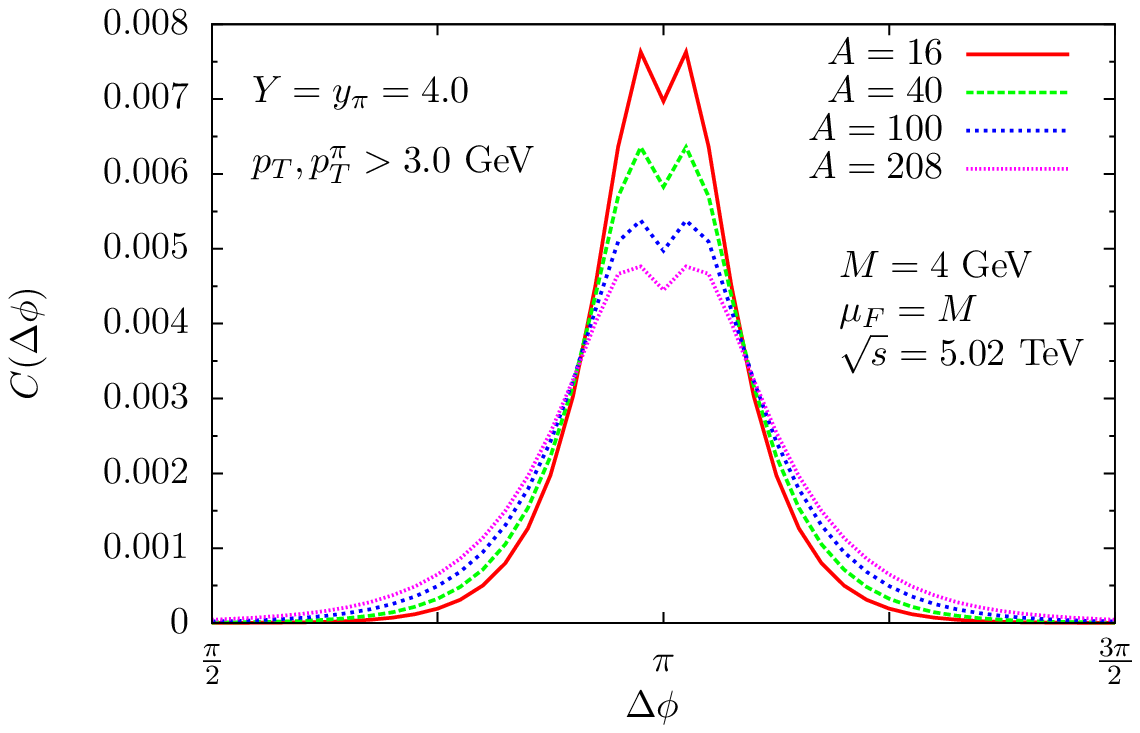}} 
\caption{(Color online) The correlation function $C(\Delta \phi)$ 
for the associated DY pair and pion production 
in $pA$ collisions at the LHC ($\sqrt{s}=5.02$ TeV) 
for different mass numbers $A$.}
\label{fig:cor1}
\end{center}
\end{figure}
\normalsize

\begin{figure}[h!]
\large
\begin{center}
\scalebox{0.7}{\includegraphics{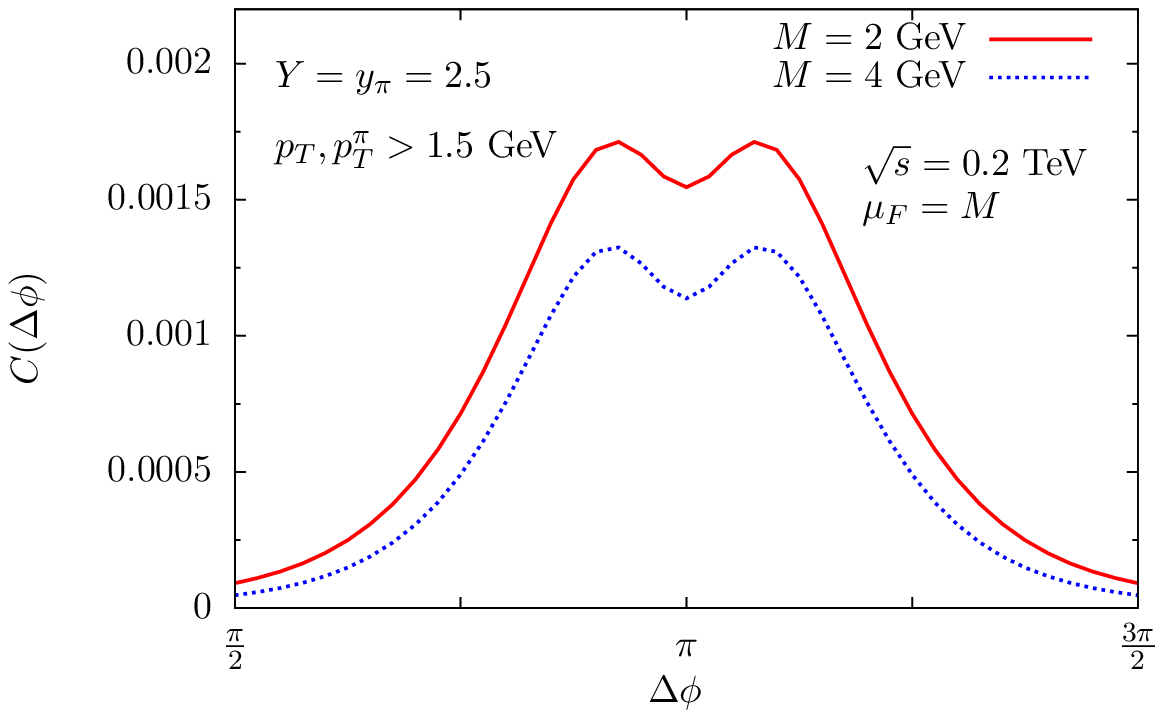}}\\ 
\scalebox{0.7}{\includegraphics{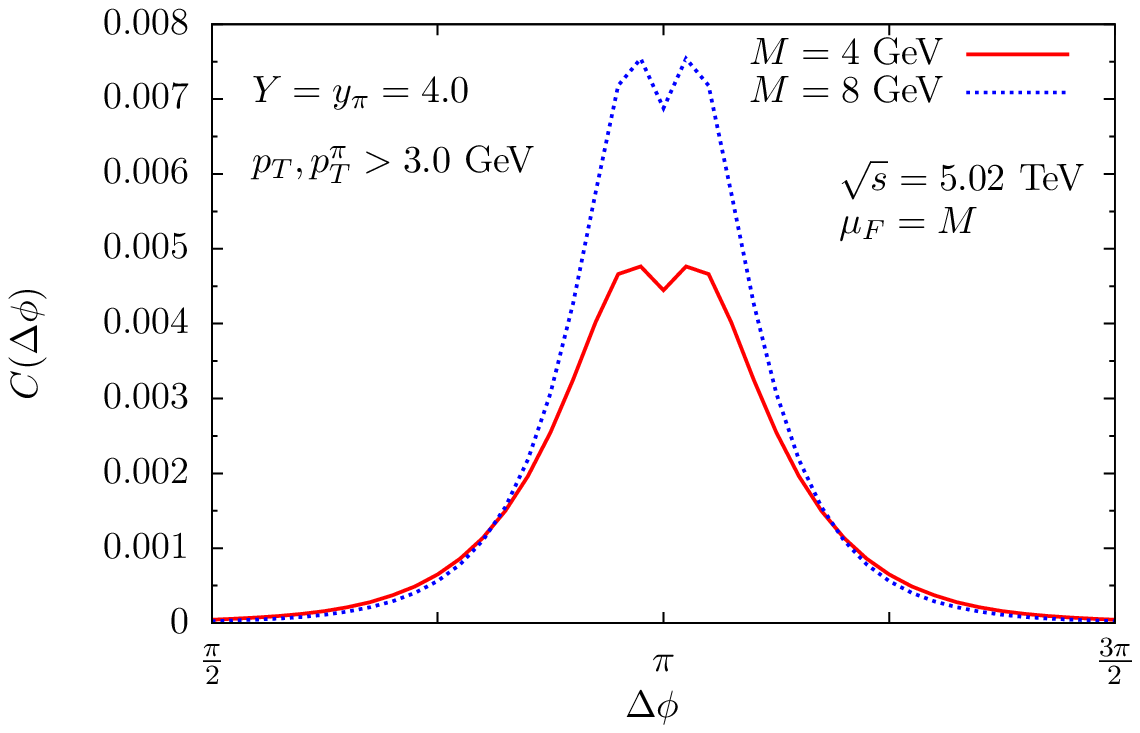}}\\
\scalebox{0.7}{\includegraphics{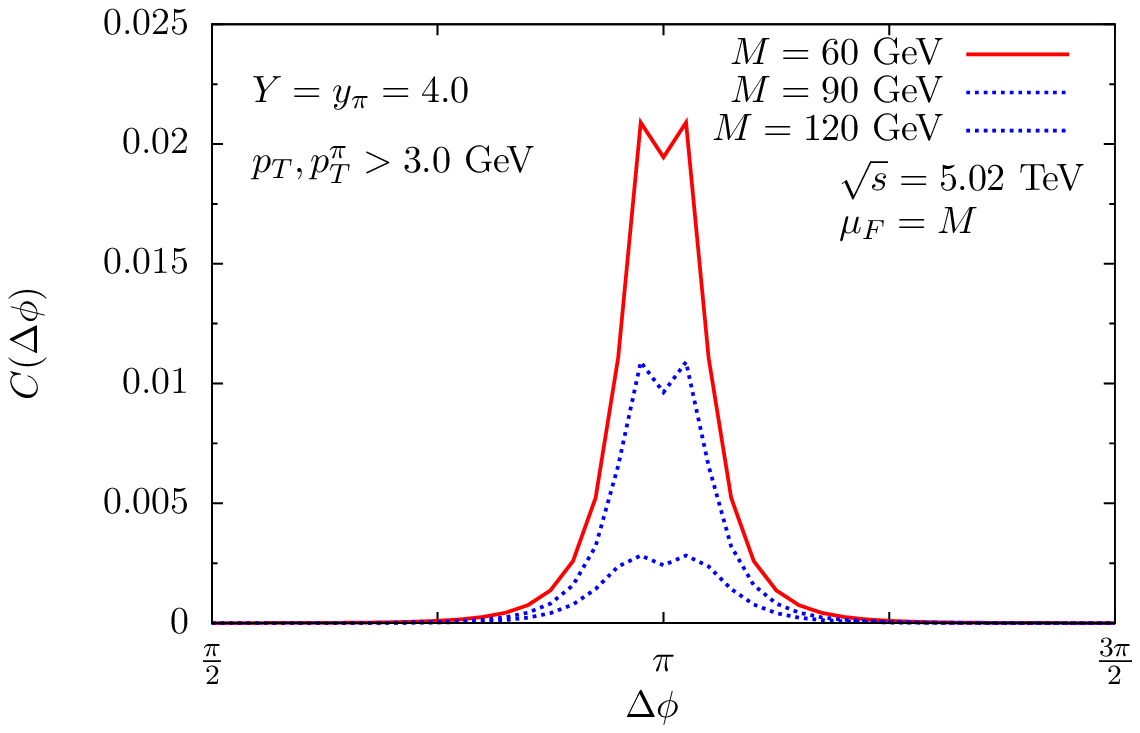}} 
\caption{(Color online) The correlation function $C(\Delta \phi)$ 
for the associated DY pair and pion production 
in $pPb$ collisions at RHIC ($\sqrt{s}=0.2$ TeV) 
and LHC ($\sqrt{s}=5.02$ TeV) energies and different values 
of the dilepton invariant mass. }
\label{fig:cor2}
\end{center}
\end{figure}
\normalsize

Finally, let us discuss the azimuthal correlation between the DY pair and a forward pion produced in $pA$ collisions 
taking into account the $Z^0$ boson contribution in addition to the virtual photon as well as the saturation effects. 
As was discussed earlier in Refs.~\cite{stastody,amir,Basso_pp}, the dilepton-hadron correlations can serve
as an efficient probe of the initial state effects. The gauge boson radiation off the projectile quark has a back-to-back 
correlation in the transverse momentum. However, the multiple scatterings of the quark in a high density gluonic system 
implies that it acquires a transverse momentum comparable with the saturation scale. As a consequence, the intrinsic 
angular correlations are expected to be suppressed, with the suppression being directly related to the magnitude of 
the saturation scale. As the saturation scale is strongly dependent on the nuclear atomic number, we expect that 
the effects predicted in Ref. \cite{Basso_pp} for the DY process in  $pp$ collisions to be amplified in the $pA$ case.
Considering the $G^*=\gamma^*/Z_0$ boson as a trigger particle, the corresponding correlation function can be written as
%
\begin{eqnarray}
C(\Delta \phi) = 
\frac{ 2\pi\, \int_{p_T, p_T^h > p_T^{\rm cut}} dp_T p_T \; 
dp_T^h p_T^h \; 
\frac{d \sigma(pA \to h G^* X)}{d Y d y_h d^2p_T d^2p_T^h }} 
{\int_{p_T > p_T^{\rm cut}} dp_T p_T \; 
\frac{d \sigma(pA\rightarrow G^* X)}{dY d^2 p_T} }\,,
\label{corr}
\end{eqnarray}
%
where $p_T^{\rm cut}$ is the experimental low cut-off on transverse momenta of the resolved $G^*$ (or dilepton) 
and a hadron $h$, $\Delta \phi$ is the angle between them. The differential cross sections entering the numerator and denominator 
of $C(\Delta \phi)$ have been derived for $pp$ collisions in Ref.~\cite{Basso_pp} taking into account both the $\gamma^*$ and $Z^0$ boson 
contributions and can now be directly generalised for $pA$ collisions by accounting the nuclear dependence of the saturation scale. 
We refer to Ref.~\cite{Basso_pp} for details of the differential cross sections. The main input in the calculation of the correlation function 
is the unintegrated gluon distribution $F(x_g,k^g_T)$, where $x_g$ and $k^g_T$ are the momentum fraction and transverse momentum 
of the target gluon, which is directly associated to the description of the QCD dynamics in the high energy limit \cite{hdqcd}. As demonstrated 
in Ref.~\cite{stastody}, the correlation function for $\Delta \phi \approx \pi$ is determined by the low-$k^g_T$ behaviour of the unintegrated 
gluon distribution which is strongly associated with the saturation effects. Since in this regime the current parametrizations for $F(x_g,k^g_T)$ 
are similar, the resulting predictions for $C(\Delta \phi \approx \pi)$ are almost model independent. In order to compare our predictions with 
those presented in Refs. \cite{stastody,Basso_pp}, in what follows we study the correlation function $C(\Delta \phi)$ taking 
the unintegrated gluon distribution (UGDF) in the following form
%
\begin{equation}
F(x_g,k^g_T) = \frac{1}{\pi Q_{s,A}^2(x_g) }\, e^{-{k^g_T}^2/Q_{s,A}^2(x_g)} \,,
\end{equation} 
%
where  
$Q_{s,A}^2 (x) = A^{1/3} c(b)\,Q_{s,p}^2(x)$ 
is the saturation scale and $Q_{s,p}^2(x)$ is given by Eq. (\ref{satsca1}). 
In numerical analysis, the CT10 NLO parametrization \cite{ct10} 
for the parton distributions and the Kniehl-Kramer-Potter (KKP) 
fragmentation function $D_{h/f}(z_h,\mu_F^2)$ of a quark 
to a neutral pion \cite{kkp} have been used. Moreover, we assume 
that the minimal transverse momentum ($p_T^{\rm cut}$) 
of the gauge boson $G^*$ and the pion $h=\pi$ in Eq.~(\ref{corr}) 
are the same and equal to 1.5 and 3.0 GeV for RHIC and LHC energies, respectively. 
As in our previous study \cite{Basso_pp}, we assume 
that the factorisation scale is given by the dilepton 
invariant mass, i.e. $\mu_F = M_{l\bar{l}}$.

The analysis of the correlation function in $pp$ collisions performed in Ref.~\cite{Basso_pp} has demonstrated that an increase of 
the saturation scale at large rapidities implies a larger value for the transverse momentum carried by the low-$x$ gluons in the target 
which generates the decorrelation between the back-to-back jets. Since the magnitude of the saturation scale is amplified by 
the factor $A^{1/3}$ in nuclear collisions we should also expect a similar effect in $pA$ collisions. In particular, the double-peak 
structure of $C(\Delta \phi)$ in the away-side dilepton-pion angular correlation function predicted to be present in $pp$ collisions \cite{Basso_pp} 
should also occur in the $pA$ case. As discussed in detail in Refs.~\cite{stastody,amir,Basso_pp}, this double peak in the region 
where $\Delta \phi \approx \pi$ is directly associated to the interplay between the local minimum of the $h+G$ cross section 
for gluon $k^g_T = |\vec{p}_T + \vec{p}_{Tq}| \rightarrow 0$, where $\vec{p}_T$ ($\vec{p}_{Tq}$) is the transverse momentum of the gauge boson (quark), 
and the two maxima for the cross section when $k_T \rightarrow Q_s$. Therefore, the double-peak structure is sensitive to the magnitude 
of the saturation scale as well. In Fig.~\ref{fig:cor1} we present our predictions for the correlation function $C(\Delta \phi)$ of the associated DY pair and pion 
in $pA$ collisions at LHC energies considering different values of the atomic mass number. We notice that the larger values of $A$ imply the stronger 
smearing of the back-to-back scattering pattern and suppress the away-side peak in the $\Delta \phi$ distribution. This behaviour is expected since 
in high energy collisions the produced parton on average has intrinsic transverse momentum of the order of the saturation scale which increases for larger $A$. 
Such an increase in $Q_s$ washes away the intrinsic back-to-back correlations. Moreover, at larger $Q_s$ one observes that the single-particle inclusive cross section 
in the denominator of Eq.~(\ref{corr}) is enhanced while the two-particle correlated cross section (in numerator of  Eq.~(\ref{corr})) is suppressed. 
As a consequence, $C(\Delta \phi)$ decreases with an increase of the saturation scale.

Our predictions for RHIC and LHC energies and $pPb$ collisions are presented in Fig.~\ref{fig:cor2} considering small and large dilepton invariant masses. 
Our results for small invariant masses, shown in the upper and middle panels, agree with those presented in Refs.~\cite{stastody,Basso_pp}. On the other hand, 
our predictions for the correlation function for large invariant masses (lower panel) are in variance with the results obtained in Ref.~\cite{Basso_pp} for $pp$ collisions. 
We also predict a double-peak structure for large invariant masses in $pPb$ collisions. As discussed before, in $pA$ collisions the saturation scale is amplified 
by a factor $A^{1/3}$, implying larger values for the average transverse momentum acquired by the quark in its multiple scatterings off the target. Moreover, 
the typical transverse momentum of the produced particles in $pA$ collisions at $\sqrt{s} = 5.02$ TeV is smaller that in $pp$ collisions at $\sqrt{s} = 14$ TeV. 
As a consequence, the impact of gluonic interactions in the produced quark is larger in $pPb$ than in $pp$ collisions. It implies a certain imbalance of the back-to-back 
photon-quark jets also for large invariant masses in $pA$ collisions, washing out the intrinsic correlations and thus generating the double-peak structure observed 
in Fig.~\ref{fig:cor2}. The away-side peak is strongly suppressed at forward rapidities and the double-peak structure is present in the kinematic range probed 
by RHIC and LHC. Consequently, we believe that our predictions can be compared with the future experimental analysis. If experimentally confirmed, this decorrelation 
and the double-peak structure are important probes of underlying saturation physics. 

A more elaborate study of the double-peak structure in the correlation function requires a multidimentional numerical analysis of the $C(\Delta \phi)$ function 
at different pion and dilepton rapidities, transverse momenta and dilepton invariant mass as well as experimental cuts. Besides, it would be instructive to investigate 
how the ISI, GS and coherence effects influence this function in various models for unintegrated gluon distributions. These questions are a subject of a separate 
big project which can be planned for the future provided that the corresponding experimental data become available.

\section{Summary}
\label{conc}

In this paper, we carried out an extensive phenomenological analysis of the inclusive DY $\gamma^*/Z^0 \to l\bar l$ process 
in $pA$ collisions within the color dipole approach. In particular, the inclusion of the $Z^0$ contribution enabled us to study 
for the first time the impact of the nuclear effects at large invariant dilepton masses. In distinction to hadron production, 
the DY reaction in $pA$ collisions is a very effective tool for study of nuclear effects since no final state interactions 
are expected, either the energy loss or absorption. For this reason, the DY process represents a direct and clean probe 
of the initial-state medium effects, not only in $pA$ interactions but also in heavy ion collisions. 

The analysis of the DY process off nuclei in different kinematic regions allows us to investigate the magnitude of particular nuclear effects.
In this paper, the contribution of the saturation, gluon shadowing (GS) and initial state energy loss (ISI) effects in DY observables 
were estimated considering the kinematical range probed at RHIC and LHC. The corresponding predictions for the dilepton invariant mass and 
transverse momentum differential distributions have been compared with available data at the LHC and a reasonable
agreement was found. Moreover, the invariant mass, rapidity and transverse momentum dependencies of the nucleus-to-nucleon 
ratio of production cross sections, $R_{pA} = \sigma^{\rm DY}_{pA}/(A \cdot \sigma^{\rm DY}_{pp})$, were estimated. 

Our results demonstrated that the ratio $R_{pA}$ is strongly modified by the GS and ISI effects. In particular, we found that both GS and ISI effects 
cause a significant suppression in DY production. While the GS effects dominate at small Bjorken-$x$ in the target, the ISI effects 
(in accordance with Eq.~(\ref{eq-ISI})) become effective at large transverse momenta $p_T$ and invariant masses $M_{l\bar{l}}$
of dilepton pairs as well as at large Feynman $x_F$ (or forward rapidities). Consequently, at forward rapidities in some kinematic regions at the LHC
one can investigate only a mixing of both (GS and ISI) effects even at large $p_T$- values. In contrast to other inclusive processes, the advantage 
of the DY reaction is due to elimination of the GS-ISI mixing by reduction of coherence effects at larger values of the dilepton invariant mass. 
Then, an investigation of nuclear suppression at large $p_T$ represents a clear manifestation of net ISI effects even at forward rapidities as 
is demonstrated in Fig.~\ref{fig:pt_lhc_large}. Thus, such a study of nuclear suppression at large dilepton invariant masses, transverse momenta and rapidities
especially at the LHC energy favours the DY process as an effective tool for investigation of the ISI effects. 

Besides, we have analysed the correlation function $C(\Delta \phi)$ in azimuthal angle $\Delta \phi$ between the produced dilepton and a forward pion 
which results by a fragmentation from a projectile quark radiating the virtual gauge boson. The corresponding observable has been studied at various energies 
in $pA$ collisions in both the low and high dilepton invariant mass ranges as well as at different rapidities of final states. We found a characteristic 
double-peak structure of the correlation function around $\Delta \phi \simeq \pi$ at various dilepton mass values and for a very forward pion. 
Our results indicated that a measurement of the correlation function at different energies at RHIC and LHC can be useful to probe underlying 
dynamics by setting further even stronger constraints on saturation physics. Finally, our results have demonstrated that the study of the DY reaction 
in $pA$ collisions is ideal to probe the nuclear effects expected to be present at high energies and large nuclei.

\section*{Acknowledgements}

E.B. is supported by CAPES and CNPq (Brazil), contract numbers 2362/13-9 
and 150674/2015-5. V.P.G. has been supported by CNPq, CAPES and FAPERGS, Brazil.
R.P. is supported by the Swedish Research Council, contract number 621-2013-428.
J.N. and M.K. are partially supported by the grant 13-20841S of the Czech Science Foundation 
(GA\v CR) and by the Grant M\v SMT LG15001. J.N. is supported by the Slovak Research and 
Development Agency APVV-0050-11 and by the Slovak Funding Agency, Grant 2/0020/14. 

\bibliographystyle{unsrt}

\end{document}